\documentstyle[emulateapj,epsfig]{article}
\renewcommand{\baselinestretch}{1.0}
\newcommand\hi{\hbox{H\kern 0.1em{\sc i}}}
\newcommand\fsi{${\mit\Sigma}$}

\lefthead{Hibbard \& Sansom}
\righthead{\hi\ in fine-structure ellipticals}
\slugcomment{To Appear in the February 2003 AJ}

\begin{document}
\title{A Search for \hi\ in five ellipticals with fine structure}

\author{J.~E. Hibbard}
\affil{National Radio Astronomy Observatory, 520 Edgemont 
Road, Charlottesville, VA, 22903; jhibbard@nrao.edu}

\bigskip
\author{A.~E. Sansom}

\affil{Centre for Astrophysics, University of Central Lancashire, 
Preston PR1 2HE, UK; AESansom@uclan.ac.uk}

\begin{abstract}

We report on VLA \hi\ spectral line observations of five early-type
galaxies classified as optically peculiar due to the presence of jets,
ripples or other optical ``fine structure''.  We detect \hi\ within
the primary beam (30$'$ HPBW) in four of the five systems. However, in
only one case is this gas associated with the targeted elliptical
galaxy. In the other cases the \hi\ is associated with a nearby
gas-rich disk or dwarf galaxy.  The one \hi\ detection is for NGC
7626, where we tentatively detect an \hi\ cloud lying between 20 and
40 kpc southwest of the galaxy center. Its origin is unclear. Our
failure to detect obvious tidal \hi\ features suggests that if these
fine-structure ellipticals are remnants of disk galaxy mergers, 
either the progenitors were gas poor, or they 
are well evolved and any gaseous tidal features have dispersed and/or
been converted into other phases. Our targeted systems all reside in
groups or clusters, and it seems likely that tidal \hi\ is shorter
lived in these environments than suggested by studies of more isolated
merger remnants.

\end{abstract}

\keywords{
galaxies: individual (NGC 3610, NGC 3640, NGC 4382, NGC 5322, NGC 7626) 
--- galaxies: interactions 
--- galaxies: ISM 
--- galaxies: peculiar 
}

\section{Introduction}
\label{sec:intro}

It was once believed that elliptical galaxies were evolved objects
with simple morphologies. However, early image processing techniques
revealed that anywhere from one-forth to one-third of such galaxies
exhibit faint morphological peculiarities in the form of shells,
ripples, and extended plumes (Malin 1978, 1979; Malin \& Carter 1980,
1983; Schweizer \& Ford 1984; Schweizer \& Seitzer 1988, 1992
hereafter SS92; Reid, Boisson \& Sansom 1994).  Numerical work soon
demonstrated that such features could be reproduced either by small
accretion events (Quinn 1984; Dupraz \& Combes 1986; Hernquist
\& Quinn 1988, 1989) or by major mergers (Hernquist \& Spergel 1992;
Hibbard \& Mihos 1995).  The appearance of such features in heretofore
``normal'' ellipticals is seen as strong support for the Toomre merger
hypothesis that (some) elliptical galaxies evolve from the merger of
disk galaxies (Toomre \& Toomre 1972, Toomre 1977).

In an effort to quantify the frequency and quantity of such
morphological peculiarities, Schweizer et al.\ (1990) introduced a
``fine-structure index'', \fsi, which increases with the amount of
morphological peculiarities (for detailed review, see Schweizer
1998). Statistically, galaxies with larger values of \fsi\ have
observational characteristics (colors and spectral-line strengths)
consistent with the presence of younger stellar populations,
supporting the idea that they represent more recent mergers in which
starbursts have occurred (Schweizer et al.\ 1990; SS92). This
association suggests that the fine-structure index might provide a
means to locate evolved merger remnants lying in the so-called ``King
Gap'' between $\sim 1$ Gyr old remnants and old ellipticals (I.\ King,
quoted in Toomre 1977).

However, some recent studies have called into question the
identification of ellipticals with large amounts of fine structure as
aged merger remnants (Silva \& Bothun 1998a, 1998b). Since such subtle
morphological peculiarities may arise from less structurally damaging
events, such as accretions, they are not unambiguous signatures of
major mergers. To remove this ambiguity, it is desirable to find a 
more direct link between a high fine-structure index and a merger
origin. 

Some clues as to the expected appearance of ``King Gap'' objects are
provided by observations of nearby systems which are undeniably the
remnants of disk-disk mergers. The best examples of such objects
are the last two members of the ``Toomre Sequence'' (Toomre 1977), a
proposed evolutionary sequence of disk-disk mergers. These two
systems, NGC 3921 and NGC 7252, are evolved merger remnants with
single nuclei, post-burst stellar populations, relaxed $r^{1/4}$ light
profiles, and with dynamical ages of $\sim$0.5--1 Gyr (Schweizer 1982,
1996, Schweizer et al.\ 1996, Hibbard \& Mihos 1995, Miller et al.\
1997, Hibbard \& Yun 1999). \hi\ mapping of these systems with the
VLA\footnote{The VLA of the National Radio Astronomy Observatory is
operated by Associated Universities, Inc., under cooperative agreement
with the National Science Foundation.} shows their tidal tails to be
rich in atomic hydrogen (Hibbard et al.\ 1994, Hibbard \& van Gorkom
1996) firmly establishing their status as disk-merger products.

Both remnants are also found to have a very low central \hi\
content. This result is somewhat surprising, not only in view of large
amounts of \hi\ which must have been sent into the central regions
during the early stages of merging (e.g. Barnes \& Hernquist 1991),
but also in view of the atomic gas-rich material that is observed
streaming back into the remnant bodies from the tidal tails (Hibbard
\& Mihos 1995, Yun \& Hibbard 2001). These studies suggest that as 
the merger evolves, the remnant should have a decreasing amount of 
fine structure, decreasing amounts of both inner and outer tidal gas, 
and a multi-epoch aging stellar population.

The present study is an attempt to locate products of gas-rich mergers
intermediate in age between recent merger remnants like NGC 3921 \&
NGC 7252 and ellipticals with less obvious peculiarities.  \hi\
mapping of morphologically peculiar early type galaxies provides
strong evidence pointing to a gas-rich merger origin for a number of
objects (e.g. Schiminovich et al. 1994, 1995; Lim \& Ho 1999; Balcells
et al. 2001; Chang et al. 2001). With this in mind, we have targeted
five ellipticals with varying amounts of optical fine structure (\fsi
= 2.0 -- 7.6) for \hi\ mapping observations with the VLA in its most
compact configuration (D-array) to look for the remnants of gas-rich
tidal features in the outer regions.  The five systems are NGC~3610,
NGC~3640, NGC~4382, NGC~5322 and NGC~7626.  In this paper we report
the results of the \hi\ observations of these five galaxies.

The present paper is structured as follows.  In \S\ref{sec:Sample} we
describe the selection criterion used to define the targets for this study.
\S\ref{sec:VLA} we describe the VLA observations and data reduction.  In
\S\ref{sec:results} we describe the results for each of the five
systems, and in \S\ref{sec:disc} we discuss these results.  Finally we
summarize our conclusions in \S\ref{sec:summary}.

\section{Sample Selection}
\label{sec:Sample}

The galaxies for this study were chosen from the sample of 69 field
elliptical and S0 galaxies for which the fine-structure index has been
tabulated by SS92. The fine-structure index is
defined as\ \ ${\mit\Sigma} = S + \log(1 + n) + J + B + X$, where $S$
is a visual estimate of the strength of the most prominent ripples,
$n$ is the number of detected ripples, $J$ is the number of luminous
plumes or ``jets'', $B$ is a visual estimate of the maximum boxiness
of isophotes, and $X$ indicates the absence or presence of an
X-structure.  The well known young ($\sim$0.5--1 Gyr) merger remnants
NGC 3921 and NGC 7252 have \fsi= 8.8 and 10.1, respectively, while the
sample of field S0's and Ellipticals studied by SS92
have values in the range \fsi\ = 0 -- 7.6.

The five galaxies chosen for this study are listed in
Table~\ref{tab:VLAobs}, along with their basic properties, group
membership, fine-structure index, and observational details.  We
selected the three galaxies in the list of SS92 with the highest
values of \fsi\ (NGC 3610, NGC 3640, NGC 4382).  We originally chose
galaxies to also cover a range of X-ray excess ($L_X/L_B$), so that we
might address whether merger remnants ``grow'' X-ray halos as tidal
gas falls back towards the remnant (Hibbard et al.\ 1994, Fabbiano \&
Schweizer 1995, Read \& Ponman 1998).  However, a re-analysis of the
ROSAT X-ray data for these systems (Sansom, Hibbard \& Schweizer 2000)
shows that all but NGC 7626 are in the lowest X-ray class, so we will
not address this question here. In an earlier study (Sansom, Hibbard \&
Schweizer 2000), we used the present sample along with the early-type
galaxies in the list of SS92 that have existing X-ray observations, to
test the hypothesis that X-ray halos form as cold tidal gas falls back
into the remnant bodies. We found no clear trends, other than the
previously reported trend for recent merger remnants to be X-ray faint
(Fabbiano \& Schweizer 1995; Georgakakis, Forbes \& Norris 2000; 
O'Sullivan, Forbes \& Ponman 2001).

\section{VLA Data}
\label{sec:VLA}

The neutral hydrogen data were collected in December of 1997 using the
VLA in its spectral line mapping mode.  The array was in its most
compact D-array configuration, providing maximum sensitivity to
extended emission. Five galaxies and several calibrators were
observed. Table~\ref{tab:VLAobs} gives a log of these observations.

Since the atomic gas within tidal debris typically has narrow
linewidths ($\sigma_v \sim$5---10 km s$^{-1}$; Hibbard et al. 1994,
Hibbard \& van Gorkom 1996), the correlator mode was chosen to give
the smallest channel spacing while still covering the expected range
of velocities for tidal gas. We therefore chose the 2AC mode with
on-line Hanning smoothing and a 3.125 MHz bandwidth, resulting in 63
spectral channels each of width $\sim$ 10.4 km s$^{-1}$ and a total
velocity coverage of $\sim650$ km s$^{-1}$. This is sufficient to
cover the spread of velocities observed in other peculiar ellipticals
($\sim 500$ km s$^{-1}$, e.g. Schiminovich et al. 1994, 1995, 2001).
However, most of the target ellipticals are members of loose groups or
clusters (cf. \S\ref{sec:results}) and the present observations do not
cover the velocity range of all group/cluster members. The
field-of-view is set by the 25m diameter of the individual array
elements, and amounts to a half-power beam width (HPBW) of 30$'$. Our
on-source integration time of $\sim$ 2 hours per source provides an
rms noise of $\sim 0.6$ mJy/beam, corresponding to a 3$\sigma$ column
density sensitivity of $\sim 7\times 10^{18}\, {\rm cm}^{-2}$ per
channel over a typical beam size of $60'' \times 50''$ (see Table
\ref{tab:VLAobs}).


The data were reduced using the AIPS software. For each observation
the absolute flux level was set by observing a standard VLA flux
calibrator.  The antenna phases were calibrated with observations of
an unresolved continuum source before and after each galaxy
observation. The bandpass was calibrated using either the flux or
phase calibrator, whichever provided the highest signal to noise.  For
details on VLA observation and calibration techniques, see Taylor,
Carilli \& Perley (1999). For the specifics of spectral line data
reduction and imaging, see Rupen (1999).

The spectral response of the VLA resulted in 55 usable channels
(channels 3--57) out of the 63 channel datacubes. The continuum was
removed using an iterative procedure which involved subtracting linear
fits to the visibilities at either end of the passband. Various
combinations of channels were used in the fit, and the resulting
datacube was mapped and reviewed to determine the range of line-free
channels.  In cases where \hi\ was detected from a companion at
velocities significantly different from the central velocity setting,
separate datacubes were made using the appropriate line-free channels 
for continuum subtraction. 

The data were mapped using a ``Robust" weighting parameter of +1
(Briggs 1995), which provides good sensitivity to faint extended
emission. Continuum intensity and continuum-subtracted line maps were
made of the entire primary beam.  The line maps were cleaned using
the standard methods in AIPS. To further increase our
sensitivity, we smoothed the data first in velocity by a factor of
four, then spatially by convolving to a resolution of
90$''$.  No further detections were made in either of the
smoothed datasets and all figures presented here use the full
resolution datacubes.

Neutral hydrogen masses are calculated according to the equation
$M_{HI} = (2.36\times 10^5)\times D_{\rm Mpc}^2 \times \,
\int S_{\rm HI} dv$, where $D_{\rm Mpc}$ is the 
distance to the source in Mpc, $S_{\rm HI}$ is the continuum
subtracted \hi\ line specific intensity in Janskys, $v$ is the
velocity expressed in km s$^{-1}$. The integral is over the line,
assuming the atomic gas to be optically thin. $M_{HI}$ is then in
units of $M_{\odot}$ per beam.  Where no detections were made, six
sigma upper limits were estimated from the velocity smoothed datacube
(i.e. over an \hi\ linewith of 42 km s$^{-1}$). \hi\ masses or mass
limits for each of the targets are listed in Table~\ref{tab:results}.
Previous limits for these galaxies were typically $10^8 - 10^9$
M$_{\odot}$ from single dish observations (Roberts et al. 1991).
Limits from our VLA data are typically an order of magnitude lower.

\section{Results}
\label{sec:results}

In this section we summarize the properties of the observed galaxies
and the results of the VLA observations. Galaxy classifications are
from the RC3 (de Vaucouleurs et al.\ 1991), and distances are take
from the {\it Nearby Galaxy Catalog} (Tully 1988), which adopts $H_o$=
75 km s$^{-1}$ Mpc$^{-1}$. Group assignment was assessed using the
Loose Galaxy Groups catalog (Garcia 1993, hereafter LGG), and the 
number of group members was evaluated via the NASA Extragalactic 
Database (NED). This information is summarized in Table~\ref{tab:VLAobs}.

We review the morphological properties of each system, listing the
values for the fine-structure index and the ``heuristic merger age'',
as tabulated by SS92. The heuristic merger age was calculated by
matching a simple two-burst model of evolving stellar populations to
the observed $UBV$ colors (see SS92 for details). We also
report spectroscopically determined age estimates available 
from the literature. In most cases, these are derived from Single
Stellar Populations (SSP) model fits to four or more spectroscopically 
derived metallicity- and age-dependent line indicies (see e.g. Gonzales 
1993, Worthey 1994), and represent a luminosity-weighted mean age for the
central stellar population. We briefly describe the results of
previous \hi\ observations (culled from the compilations of Huchtmeier
\& Richter 1989 and Martin 1998). We note that our VLA observations
are most sensitive to narrow \hi\ lines, and may miss
\hi\ emission spread over a broad range of velocity.  For such
emission the large-beam single dish limits are still relevant.

Finally, we present the results of our VLA \hi\ mapping observations,
which are summarized in Table~\ref{tab:results}.  Figure
\ref{fig:momnt0} shows the integrated \hi\ line maps for each of our
five target galaxies contoured upon a greyscale representation of the
optical light, taken from the Digital Sky Surveys (DSS).  No \hi\ was
detected from within the optical bodies of any of the target
galaxies. Only NGC~7626 shows evidence of some associated \hi, out at
projected radii from $\sim 1.5 - 3$ arcmin from the galaxy centre.

\subsection{NGC 3610 = UGC 6319}
\label{sec:n3610}

NGC 3610 is a member of the NGC 3642 group (LGG 234), with five cataloged
members. The NGC 3642 group itself is a sub-group of the rich (N=170) 
group No.\ 94 of Geller \& Huchra (1983). 
The optical velocity from the Updated Zwicky Catalog 
(Falco et al.\ 1999) is 1696 km s$^{-1}$, whereas the \hi\
observations were centered at the old RC3 value of 1787 km s$^{-1}$. 
This will not effect our detection sensitivity, since this difference
is well within the observed band.

NGC 3610 has the highest value of \fsi\ of any of the early types in
the SS92 sample (\fsi=7.60, heuristic merger age 4--7 Gyr).  Only the
young merger remnants NGC 3921 and NGC 7252 have higher values (\fsi=
8.8 and 10.1, respectively). The optical peculiarities include two or
three shells, a central `X'-structure, plumes, and extraordinarily
boxy outer isophotes (Forbes \& Thomson 1992, Whitmore et al.\
1997). While the outer isophotes are boxy, the inner isophotes are
disky, due to a inner disk embedded within its spheroid (Scorza \&
Bender 1990). The system as a whole is rotationally flattened
($V/\sigma=1$; Scorza \& Bender 1990), and the inner disk is
kinematically distinct, with $V/\sigma=4.5$ (Rix \& White 1992).
There is no color differences between the main body and the inner
disk, and both seem to be comprised of stars with similar ages (Scorza
\& Bender 1990; Whitmore et al.\ 1997). These observations are seen as
evidence for an accretion, rather than a merger, origin (Scorza \&
Bender 1990, Silva \& Bothun 1998b). However, the presence of an
intermediate aged centrally concentrated globular cluster population
(Whitmore et al.\ 1997, 2002) suggests a merger origin. The age of the
red, metal rich globular cluster population is consistent with the
broadband near-IR and optical age estimates of $\sim$4--7 Gyr (Silva
\& Bothun 1998b, SS92).

Previous single dish \hi\ observations of NGC 3610 were conducted by
Bieging (1978), who set a 3-sigma upper limit of $\int S_{\rm
HI} dv < 2.4$ Jy km s$^{-1}$ using the Effelsberg 100-m telescope
(8$'$ HPBW). This places a limit on the \hi\ mass to $B-$band
luminosity of $M_{HI}/L_B < 0.012\, M_\odot L_\odot^{-1}$.  We find no
\hi\ emission within the primary beam of the VLA observations.  The
new \hi\ limit is over an order of magnitude lower than the previous
limit ($M_{HI}/L_B < 4\times10^{-4}\, M_\odot L_\odot^{-1}$).  None of
the other members of the LGG 234 group fall within the primary beam
(30$'$ HPBW, or 130 kpc radius at the adopted distance of 29.2 Mpc)
and velocity coverage of the present \hi\ observations.

\subsection{NGC 3640 = UGC 6368}
\label{sec:n3640}

NGC 3640 is a member of NGC 3640 group (LGG 233), with seven
cataloged members.  However, none of the other group members fall within
the primary beam (105 kpc radius at the adopted distance of 24.2 Mpc)
and velocity coverage of the present \hi\ observations. The E pec
galaxy NGC 3641 lies just $2\farcm9$ south of NGC 3640, but its
velocity ($V_\odot$=1755 km s$^{-1}$) lies outside the observed band.

\begin{figure*}
\plotone{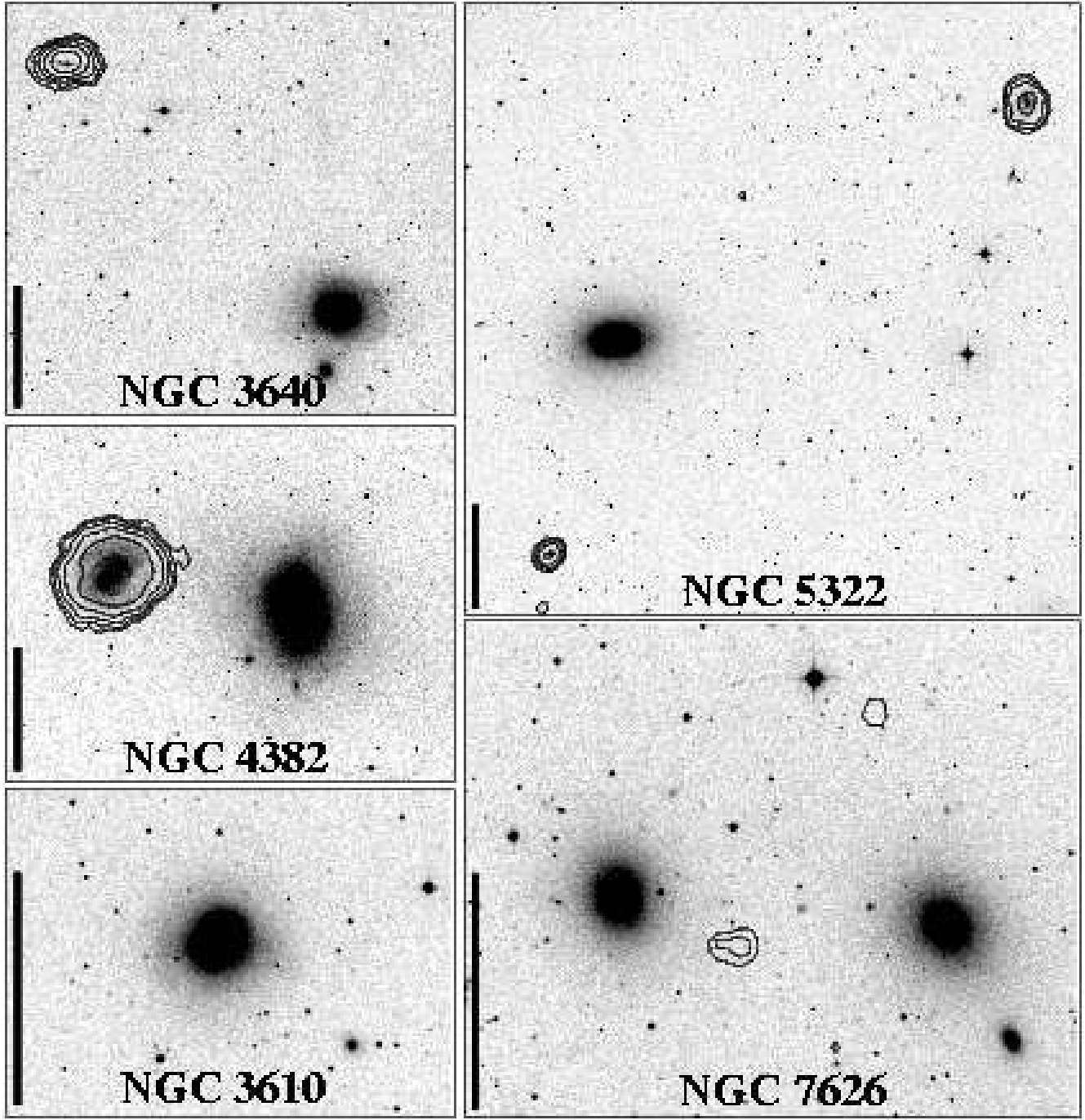}
\caption{
Montage of images for five E/SO galaxies with fine structure.
Contours of \hi\ column density are drawn upon a greyscale
representation of the second generation blue Digital Sky Survey
images. Image are oriented such that North is up and east is to the 
left. Clockwise from top left are the fields for NGC 3640, NGC 5322,
NGC 7626, NGC 3610 and NGC 4382. Starting contours are drawn at levels
of [27.6, 25.8, 24.4, 0.0, 24.5] Jy km s$^{-1}$ (respectively), which
corresponds to an \hi\ column density of $1\times 10^{19}$
cm$^{-2}$. Successive contours are drawn a factor of two higher. The
vertical bar in each frame is 5$'$ long. Of these five galaxies, only
NGC 7626 has \hi\ directly associated with the targeted elliptical
(see Table~\ref{tab:results}). In these images \hi\ is detected in: an
uncataloged companion northeast of NGC 3640; MCG+10-20-039, a southern
companion to NGC 5322; UGC 8714, a companion lying to the northwest of
NGC 5322; a cloud lying west of NGC 7626; and NGC 4394 east of NGC
4382.
}
\label{fig:momnt0}
\end{figure*}

Along with NGC 4382, NGC 3640 has the second highest value of the
fine- structure index in the SS92 compilation (\fsi=6.85, heuristic
merger age 6---8 Gyr).  It has a number of morphological
peculiarities, including shells, boxy isophotes, a minor-axis dust
lane, and like NGC 3610 it is a fast rotator ($V/\sigma$=1.5; Prugniel
et al.\ 1988).  The second generation blue DSS image shows what
appears to be a faint tidal tail extending directly to the north 
by about 10$'$ (see Fig.~\ref{fig:momnt0}).
Prugniel et al.\ (1988) argue that NGC 3640 is a
merger in progress, probably with a gas-poor system.  Both the
heuristic merger age derived by SS92 and the near-IR colors of NGC
3640 argue against a major starburst having occurred within the
previous 3 Gyr (Silva \& Bothun 1998a, 1998b).

NGC 3640 was previously searched for \hi\ by Shostak et al.\ (1975), who
set an upper limit of $\int S_{\rm HI} dv < 2.9$ Jy km s$^{-1}$ using
the NRAO 300 ft telescope (9.3$'$ HPBW), and by Lake \& Schommer
(1984), who set an upper limit of $\int S_{\rm HI} dv < 0.6$ Jy km
s$^{-1}$ using the 305-m dish at Arecibo (3.3$'$ HPBW). This places a
limit on the \hi\ mass to $B-$band luminosity of $M_{HI}/L_B < 0.003\,
M_\odot L_\odot^{-1}$.

\begin{figure*}
\plotone{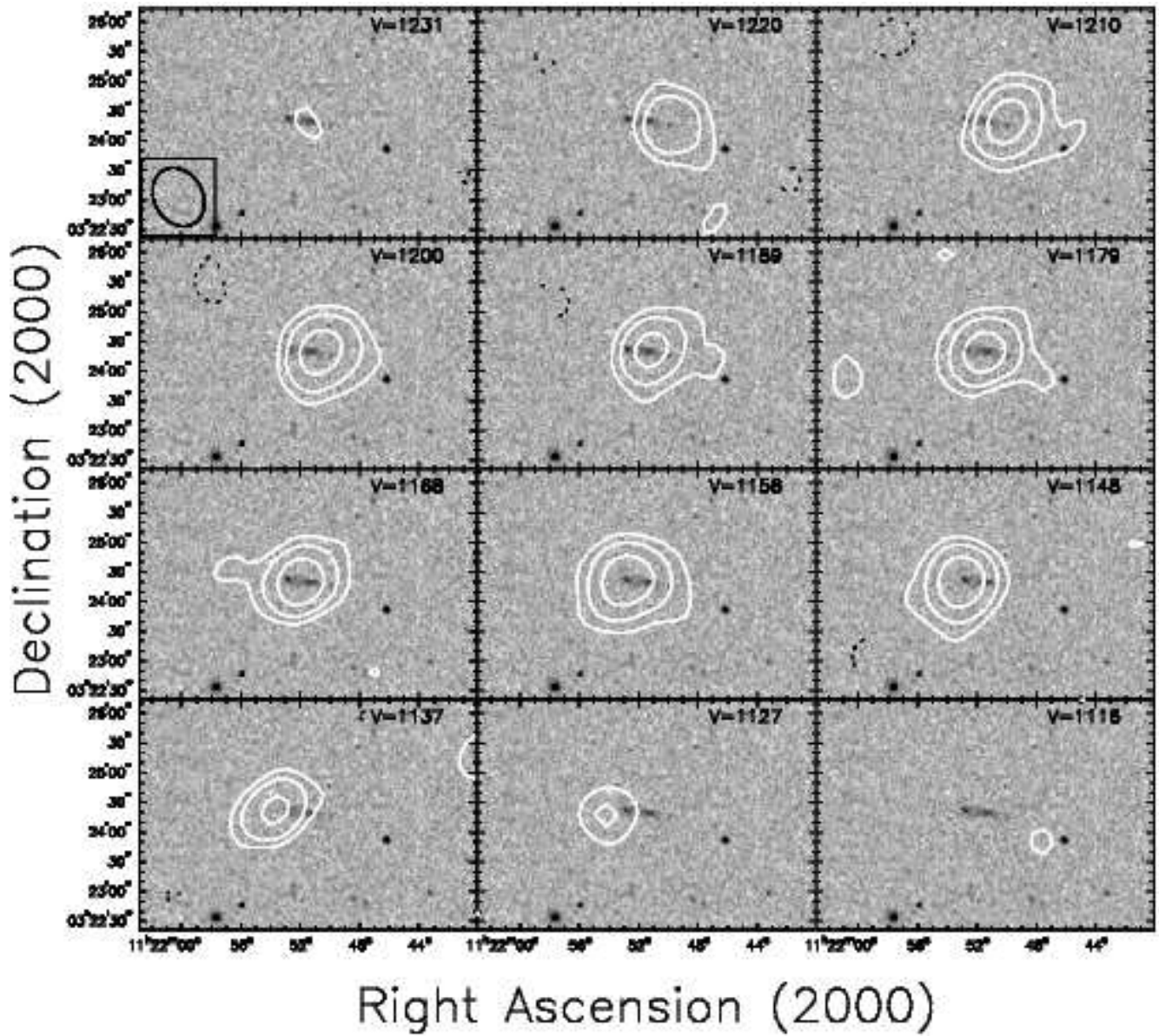}
\caption{
\hi\ channel maps contoured upon a greyscale representation of the 
DSS optical image for the \hi\ detected companion lying 15.1$'$ (105 kpc) 
to northeast of NGC 3640, at 11:21:51.5 +03:24:17 (J2000). 
The heliocentric velocity of each channel is given in the upper 
right of each panel, and the size of the synthesized beam
($62''\times49''$ FWHM) is indicated by the inscribed ellipse
in the lower left corner of the first panel.
Contours are drawn at levels of [-3, 3, 6, 12] $\times$ 
1.2 mJy beam$^{-1}$,
where the lowest contour corresponds to an \hi\ column density of
$1.4\times 10^{19}$ cm$^{-2}$.}
\label{fig:n3640ch}
\end{figure*}

We detect no \hi\ emission associated with NGC 3640 
($M_{HI}/L_B < 6\times10^{-4}\, M_\odot L_\odot^{-1}$).  While none of
the other members of the LGG 233 group fall within our VLA settings,
we do detect \hi\ in an uncataloged galaxy lying 15.1$'$ (105 kpc) to
northeast at 11:21:51.5 +03:24:17 (J2000).  The integrated \hi\
emission is contoured on the DSS image in Fig.~\ref{fig:momnt0}, and
the \hi\ channel maps of the emission are given in
Figure~\ref{fig:n3640ch}. From the DSS image, the companion appears to
be a highly-inclined low surface brightness dwarf spiral, and the \hi\
kinematics are characteristic of regular disk rotation.  The total
detected \hi\ flux, after correction by primary beam attenuation, is
2.4 Jy km s$^{-1}$, corresponding to an \hi\ mass of $3.3\times 10^8
M_\odot$, assuming the companion lies at the distance of NGC 3640
listed in Table~\ref{tab:VLAobs}.

There is an unresolved radio continuum source that lies 47$''$ to the
southeast of NGC 3640, outside its main body, but within the fainter
optical isophotes, with a flux of 41 mJy at 1.4 GHz.  There is no
optical counterpart on the DSS image, and this is probably a
background radio source.

\subsection{NGC 4382 = M85 = UGC 7508}
\label{sec:n4382}

NGC 4382 is a member of the Virgo I group (LGG 292), with 126
cataloged members.  Together with the elliptical VCC 797 (lying
2.9$'$ to the south of NGC 4382) and the barred spiral NGC
4394 (lying 7.6$'$ to the east), it forms
Redshift Survey Compact Group No.\ 54 (Barton et al.\ 1996). The dS0
galaxy IC 3292 is also likely associated (see Table~\ref{tab:results}).  
This makes this association a compact group within a
subgroup of a moderately rich cluster.  None of the other Virgo I
group members fall within the primary beam (73 kpc radius at the
adopted distance of 16.8 Mpc) and velocity coverage of the present
\hi\ observations.

Along with NGC 3640, NGC 4382 has the second highest value of the
fine-structure index in the SS92 compilation ($\Sigma$=6.85, heuristic
merger age 4--7 Gyr), with more than twelve irregular ripples,
significant isophotal twists, and a plume to the north (Burstein 1979,
Schweizer \& Seitzer 1988), and exhibiting minor-axis rotation (Fisher
1997). It has centrally enhanced $H\beta$ and Mg$_2$ absorption,
suggesting a relatively young ($<$ 3 Gyr) central population (Fisher,
Franz \& Illingworth 1996; Terlevich \& Forbes 2002, hereafter TF02).

NGC 4382 has been observed numerous times with single dish telescopes,
with the best limit obtained by Burstein, Krumm \& Salpeter (1987),
using Arecibo (3.3$'$ HPBW) and sampling both the central position of
NGC 4382 and pointings offset by 3$'$ to each of the four compass
points.  No \hi\ is detected associated with NGC 4382, resulting in
limits of $\int S_{\rm HI} dv < 0.2$ Jy km s$^{-1}$ and $M_{HI}/L_B <
3\times 10^{-4}\, M_\odot L_\odot^{-1}$. Burstein et al.\ claim a weak
narrow \hi\ feature in the pointing 3$'$ south and close to the
velocity of NGC 4382, which they tentatively associate with VCC 797.

\begin{figure*}
\plotone{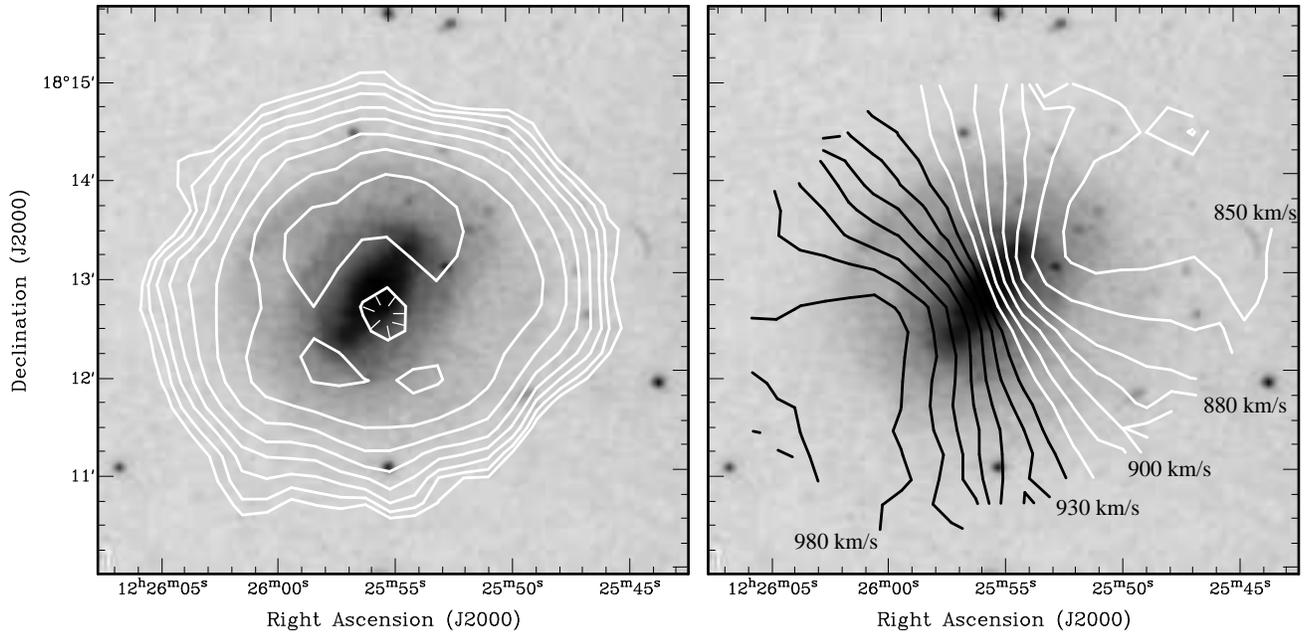}
\caption{Integrated \hi\ 
intensity (left) and 
intensity weighted isovelocity field (right) contoured
upon a greyscale representation of the 
DSS optical image of NGC 4394, the barred companion east of NGC 4382.
The integrated \hi\ intensity is contoured at 48.9 Jy km s$^{-1}$
$\times 2^{n/2}$, [n=0,1,2...], where the lowest contour
corresponds to an \hi\ column density of $2\times 10^{20}$ 
cm$^{-2}$. The hatched contour represents a central depression
in the \hi\ distribution (see Fig.~\ref{fig:n4382ch}).
The isovelocity field is contoured
from 850---990 km s$^{-1}$ in steps of 10 km s$^{-1}$, with white
contours representing blue-shifted velocities relative to 
systemic, and black contours representing red-shifted
velocities.
}
\label{fig:n4394mom01}
\end{figure*}

\begin{figure*}
\plotone{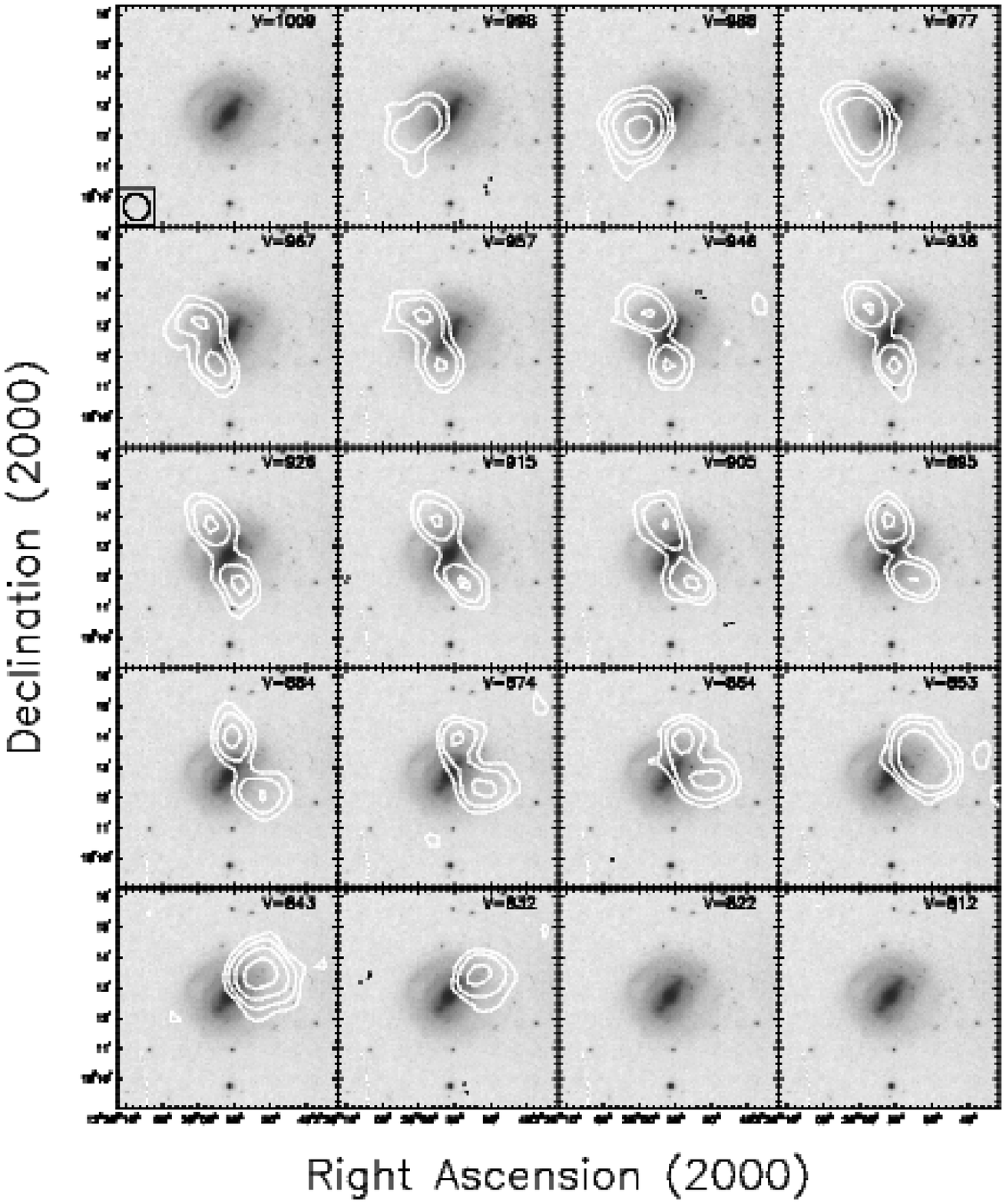}
\caption{
\hi\ channel maps contoured upon a greyscale representation of the 
DSS optical image for NGC 4394, the barred companion lying 
7.6$'$ (37 kpc) to the east of NGC 4382. 
The heliocentric velocity of each channel is given in the upper 
right of each panel, and the size of the synthesized beam
($54''\times 50''$ FWHM) is indicated by the inscribed ellipse
in the lower left corner of the first panel.
Contours are drawn at levels of [-3, 3, 6, 12, 24] $\times$ 
0.76 mJy beam$^{-1}$, where the lowest contour corresponds to 
an \hi\ column density of $9.8\times 10^{18}$ cm$^{-2}$. 
}
\label{fig:n4382ch}
\end{figure*}

We detect no \hi\ emission associated with NGC 4382, to a limit of
$\int S_{\rm HI} dv < 0.12$ Jy km s$^{-1}$, placing a limit of
$M_{HI}/L_B < 2\times 10^{-4}\, M_\odot L_\odot^{-1}$.  We do not
confirm the tentative detection of \hi\ with VCC 797, nor do we detect
the dS0 IC 3292.  We do detect $4.8\times 10^8\, M_\odot$ of
\hi\ in the barred spiral NGC 4394 (also detected by Burstein 
et al. and others).  The gas in this galaxy forms an asymmetric ring,
with the \hi\ concentrated to the spiral features outside of the bar
(Fig.~\ref{fig:n4394mom01}). The kinematics are very regular and
symmetric, as seen in the intensity weighted velocity map
(Figure~\ref{fig:n4394mom01}) and the channel maps
(Figure~\ref{fig:n4382ch}).


\subsection{NGC 5322 = UGC 8745}
\label{sec:n5322}

NGC 5322 is the dominant elliptical of the NGC 5322 group (LGG 360),
which has ten cataloged members. However, none of the other group
members fall within the primary beam (140 kpc radius at the adopted
distance of 31.6 Mpc) and velocity coverage of the present \hi\
observations (but see below).  NGC 5322 has a recently updated optical
velocity from the Updated Zwicky Catalog (Falco et al.\ 1999) of 1781
km s$^{-1}$, whereas the \hi\ observations were centered at the old
RC3 value of 1915 km s$^{-1}$. This difference is still well within
the observed band.

NGC 5322 has less obvious optical peculiarities than the other systems
in this sample ($\Sigma$=2.00, heuristic merger age 5--8 Gyr). These
include outer boxy and inner disky isophotes (Bender 1988). It also
has peculiar stellar kinematics, with a small, counter-rotating disk
embedded in a slowly rotating bulge (Bender 1988, Bender \& Surma
1992, Rix \& White 1992). There is a weak central radio source with
symmetric jets aligned perpendicular to the central disk (Feretti et
al.\ 1984), and the nuclear optical spectrum is characterized as a
LINER. The central colors and [Mg/Fe] ratio of NGC 5322 suggest the
presence of a 1--3 Gyr old central starburst population (Bender \&
Surma 1992, Silva \& Bothun 1998a, 1998b).  A luminosity weighted age of
$4.2\pm 0.6$ Gyr was estimated from fitting SSP models to 20 central
line-strengths in NGC 5322 (Proctor \& Sansom 2002), supporting the
existence of a relatively young component.

NGC 5322 was previously searched for \hi\ using the NRAO Greenbank 140
foot telescope ($21'$ HPBW) by Knapp \& Gunn (1982), obtaining an
upper limit of $\int S_{\rm HI} dv < 8.5$ Jy km s$^{-1}$, resulting in
the limit $M_{HI}/L_B < 0.022\, M_\odot L_\odot^{-1}$.  

\begin{figure*}
\plotone{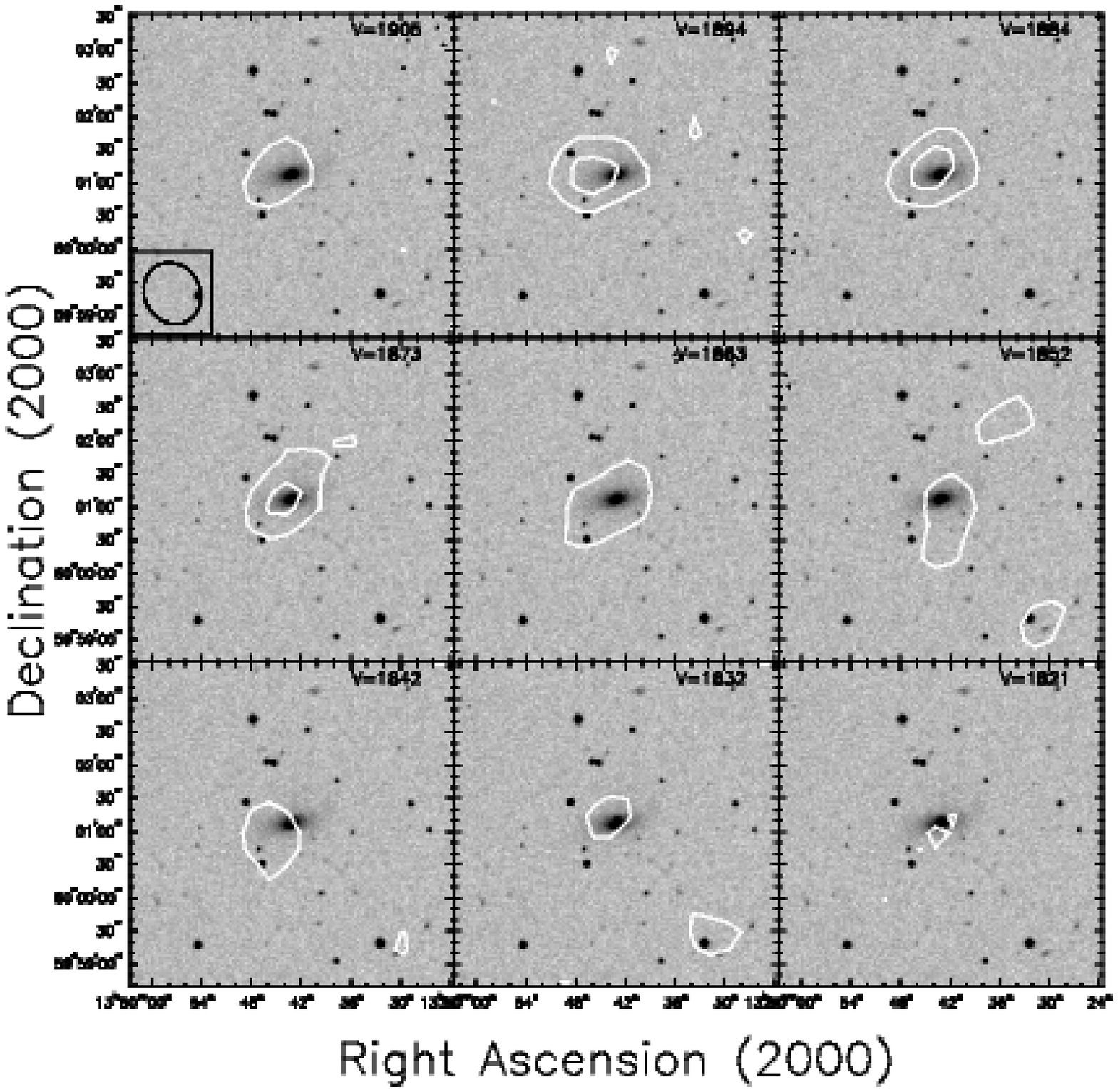}
\caption{
\hi\ channel maps contoured upon a greyscale representation of the 
DSS optical image for MCG+10-20-039, the companion lying
10.9$'$ (100 kpc) to the south of NGC 5322. 
The heliocentric velocity of each channel is given in the upper 
right of each panel, and the size of the synthesized beam
($57''\times 50''$ FWHM) is indicated by the inscribed ellipse
in the lower left corner of the first panel.
Contours are drawn at levels of [-3, 3, 6] $\times$ 
0.70 mJy beam$^{-1}$, where the lowest contour corresponds to 
an \hi\ column density of $8.5\times 10^{18}$ cm$^{-2}$. 
}
\label{fig:n5322Sch}
\end{figure*}

\begin{figure*}
\plotone{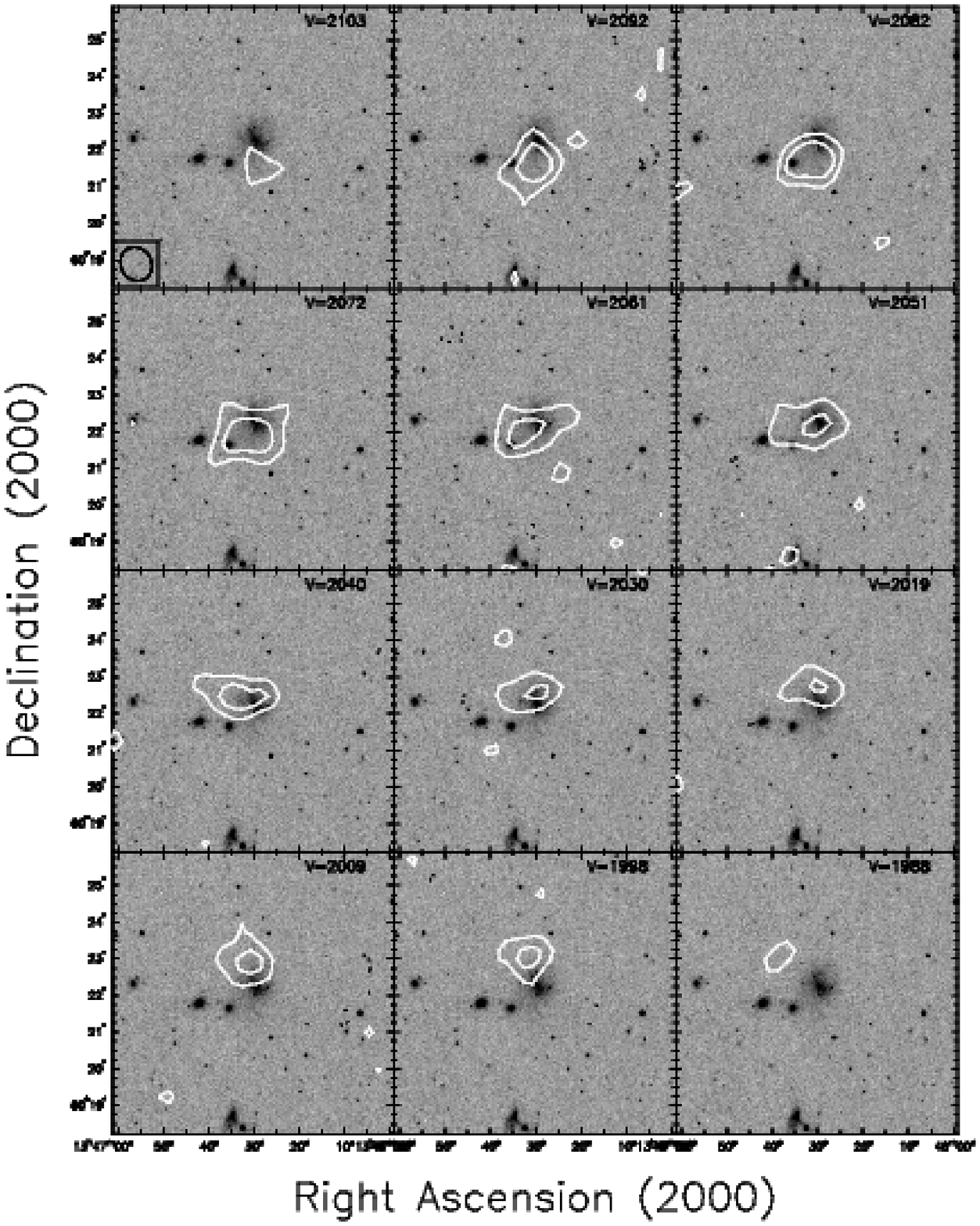}
\caption{
\hi\ channel maps contoured upon a greyscale representation of the 
DSS optical image for UGC 8714, the companion lying
23.2$'$ (215 kpc) to the northwest of NGC 5322. 
The heliocentric velocity of each channel is given in the upper 
right of each panel, and the size of the synthesized beam
($57''\times 50''$ FWHM) is indicated by the inscribed ellipse
in the lower left corner of the first panel.
Contours are drawn at levels of [-3, 3, 6] $\times$ 
2.9 mJy beam$^{-1}$, where the lowest contour corresponds to 
an \hi\ column density of $3.5\times 10^{19}$ cm$^{-2}$.
}
\label{fig:n5322NWch}
\end{figure*}

We detect no \hi\ emission associated with NGC 5322.  The new \hi\
limit is two orders of magnitude lower than the previous limit
($M_{HI}/L_B < 2\times 10^{-4}\, M_\odot L_\odot^{-1}$). We do detect
\hi\ associated with two other nearby galaxies: MCG +10-20-039, an 
early type disk system lying $10.9'$ (100 kpc) to the south, and UGC
8714, a dwarf irregular lying $23.2'$ (210 kpc) to the northwest.  MCG
+10-20-039 has no previously determined redshift, while UGC 8714 is a
known member of the NGC 5322 group. UGC 8714 was previously detected
in the \hi\ line by Schneider et al. (1992).  See Fig.~\ref{fig:momnt0} 
for the integrated \hi\ contoured on the DSS image,
Figure~\ref{fig:n5322Sch} for channel maps of MCG +10-20-039, and
Figure~\ref{fig:n5322NWch} for channel maps of UGC 8714. The \hi\
kinematics of MCG +10-20-039 do not appear very organized, but this
may be a result of our relatively coarse spatial resolution. The \hi\
kinematics of UGC 8714 are characteristic of a highly inclined
rotating disk.

\subsection{NGC 7626 = UGC 12531}
\label{sec:n7626}

NGC 7626 is a one of the two brightest ellipticals in the Pegasus I
cluster, lying just $6.9'$ (90 kpc at the adopted distance of 45.6
Mpc) and 420 km s$^{-1}$ away from the dominant elliptical galaxy of
that cluster, NGC 7619. Both galaxies are part of the Pegasus group
(LGG 473), with 25 cataloged members. However, the optical velocity of
NGC 7619 ($V_\odot$=3758 km s$^{-1}$) and most of the other cluster
members fall outside of the high-velocity end of our bandpass
($V_\odot >$ 3730 km s$^{-1}$).  

NGC 7626 has a modest amount of fine structure ($\Sigma$=2.60,
heuristic merger age 7--9 Gyr).  Its morphological peculiarities
include outer and inner shells, a major-axis dust lane, and an inner
`X'-structure (Jedrzejewski \& Schechter 1988; Forbes \& Thomson 1992;
Balcells \& Carter 1993; Forbes, Franx \& Illingworth 1995). HST
imaging reveals a symmetric dust feature within the inner 0.5$''$
(Forbes et al. 1995, Carollo et al. 1997).  This galaxy is noted for
its very strange velocity field, which exhibits core rotation about an
intermediate axis, and has shell-like regions of distinct kinematics
around the core (Jedrzejewski \& Schechter 1988, Balcells \& Carter
1993, Longhetti et al 1998). The kinematically decoupled core has an
enhanced Mg$_2$ index relative to the outer regions (Davies Sadler \&
Peletier 1993), suggesting that its formation was accompanied by rapid
star formation.  Single stellar population models of the core, inner,
and global line indicies suggest uniformly old ages for the central
stellar populations of both NGC 7626 and NGC 7619 (7--12 Gyr; Gonzalez
1993, Longhetti et al. 2000, Trager et al. 2000, TF02).

\begin{figure*}
\plotone{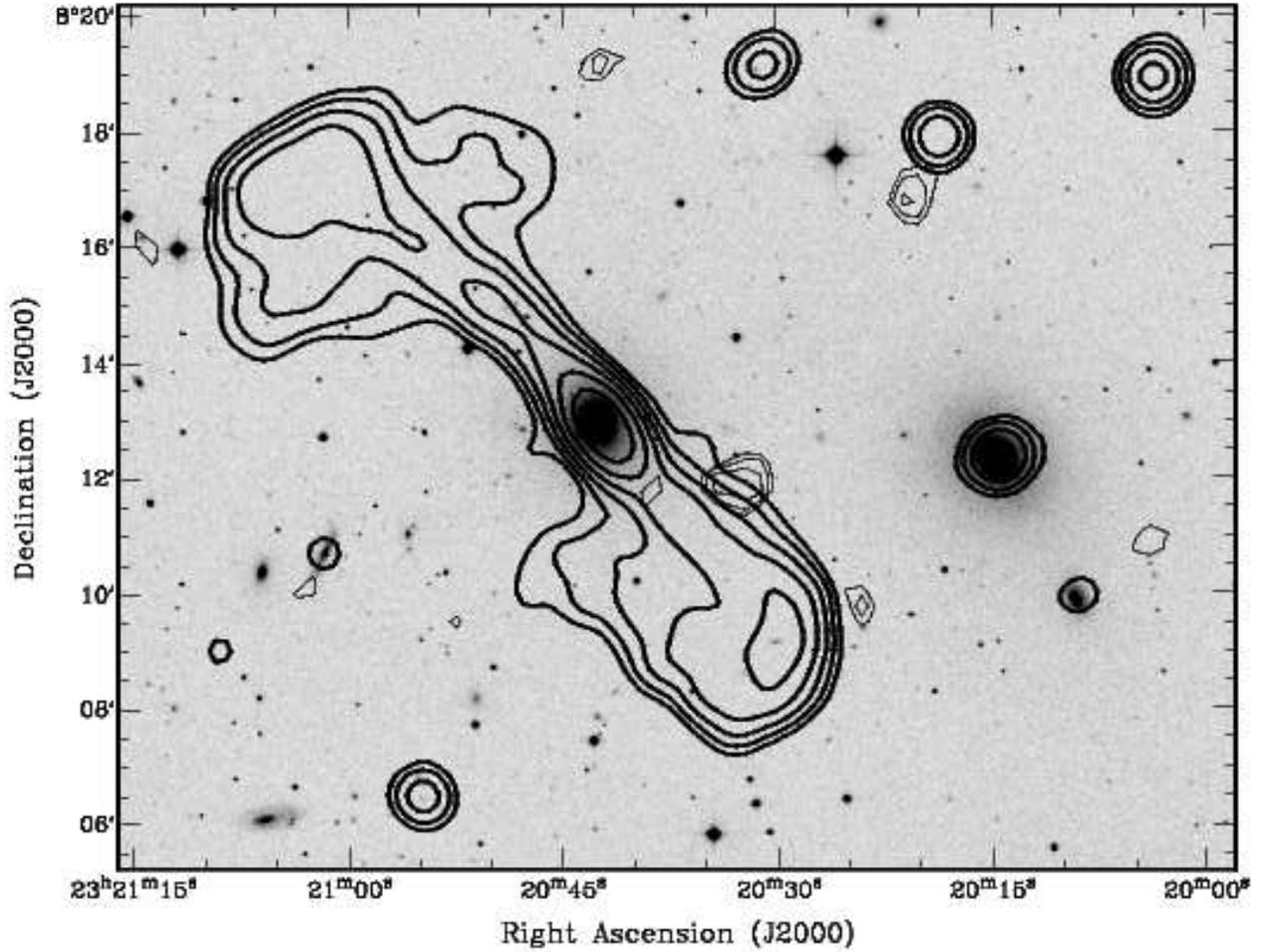}
\caption{Radio continuum and \hi\ emission
in the NGC 7626 field. Thick contours show the
distribution of the 1.4 GHz radio continuum. 
Contours start at 2 mJy beam$^{-1}$, with successive contours 
a factor of two higher than the previous contour. NGC 7619 lies
directly to the west and is also a radio source, and NGC 7631 is the
Sb galaxy lying directly to the east. Thin countours
show the \hi\ line emission integrated over the velocity range
3470---3600 km s$^{-1}$. Contours are drawn at levels of 
[1, 2, 4] $\times$ 10 mJy beam$^{-1}$ km s$^{-1}$, where the 
lowest contour corresponds to an \hi\ column density of 
$4\times 10^{18}$ cm$^{-2}$.
}
\label{fig:n7626ch0}
\end{figure*}

NGC 7626 is a radio galaxy, with two symmetric jets on either side of
a central core (Jenkins 1982, Birkinshaw \& Davies 1985). Our radio
continuum map is shown contoured on the DSS image in
Figure~\ref{fig:n7626ch0}. The lobes are edge brightened and extend
8.0$'$ (106 kpc) to the northeast and 6.3$'$ (84 kpc) to the
southwest. The core has a flux of 290 mJy at the observed sky
frequency of 1404 MHz, while the entire radio structure has a total
flux of 1.12 Jy. The radio jet is tilted by $\sim35^\circ$ with
respect to the dust lane imaged by HST (Forbes et al. 1995, Carollo et
al. 1997). NGC 7619 has an unresolved radio continuum source with a
flux of 25 mJy. NGC 7626 has a compact X-ray source at its center,
most likely associated with the central engine, while NGC 7619 is a
strong, extended X-ray source with an X-ray `tail' (Trinchieri et
al. 1997).

The region around and between NGC 7626 and NGC 7619 was mapped in the
\hi\ line by Kumar \& Thonnard (1983) with Arecibo  (3.3$'$ HPBW). No 
\hi\ was detected at any of the eleven pointings, with resulting limit of
$\int S_{\rm HI} dv < 1.1$ Jy km s$^{-1}$. A more stringent limit was
set with the Effelsberg 100-m telescope ($9.3'$ HPBW) by Huchtmeier
(1995) using a single pointing on NGC 7626, giving $\int S_{\rm HI} dv
< 0.7$ Jy km s$^{-1}$ and $M_{HI}/L_B < 0.007\, M_\odot L_\odot^{-1}$

\begin{figure*}
\plotone{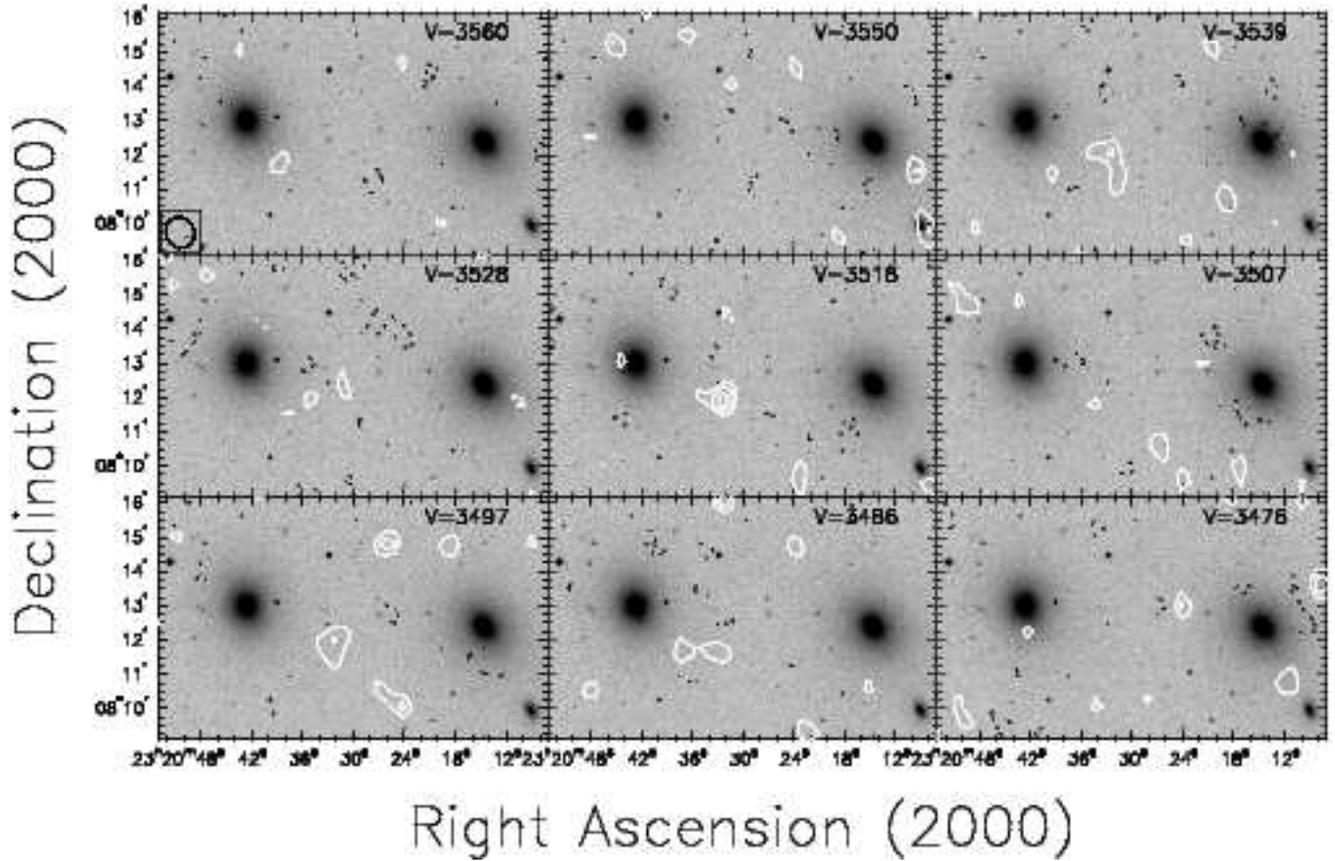}
\caption{
\hi\ channel maps contoured upon optical image from 
the DSS for NGC 7626 (east) and NGC 7619 (west).
The heliocentric velocity of each channel is given in the upper 
right of each panel, and the size of the synthesized beam
($55''\times 49''$ FWHM) is indicated by the inscribed ellipse
in the lower left corner of the first panel.
Contours are drawn at levels of [-3, -2, 2, 3, 4] $\times$ 
0.46 mJy beam$^{-1}$, where the lowest contour corresponds to 
an \hi\ column density of $4\times 10^{18}$ cm$^{-2}$.
}
\label{fig:n7626ch}
\end{figure*}

\begin{figure*}
\plotone{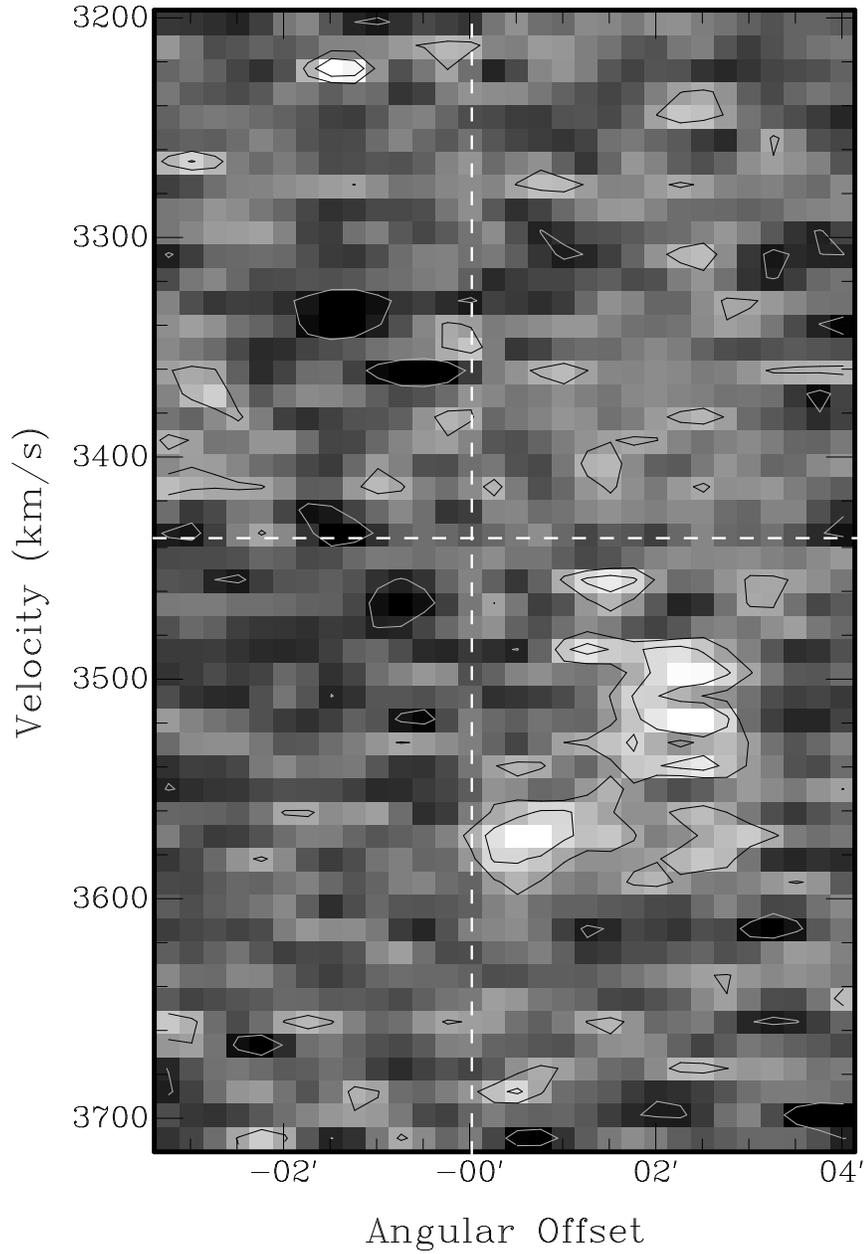}
\caption{Position-Velocity plot through the peak in the \hi\ emission
in the NGC 7626 field. The position angle of the simulated ``slit'' is
90$^\circ$ (i.e., parallel to the right ascension axis), and
centered at a declination of +08:11:42 (J2000). The right
ascension and systemic velocity of NGC 7626 are indicated by vertical
and horizontal dashed lines, respectively.
Contours are drawn at [-1, 1, 2]$\times$ 0.46 mJy beam$^{-1}$.}
\label{fig:n7626pv}
\end{figure*}

\begin{figure*}
\plotone{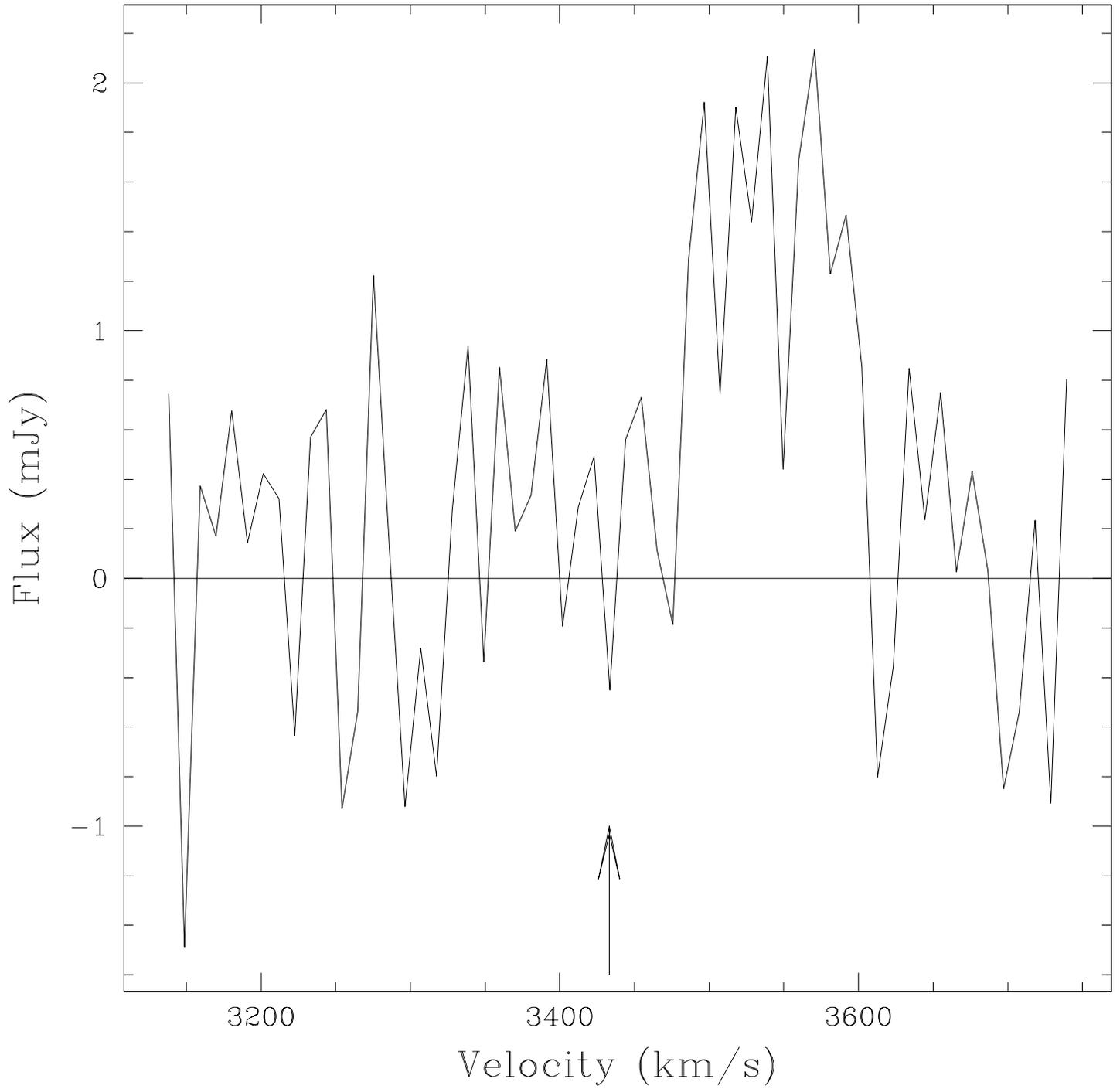}
\caption{The integrated \hi\ line profile extracted over 
the region containing the two \hi\ features in the moment map. The
velocity of NGC 7626 is indicated by the arrow.}
\label{fig:n7626sp}
\end{figure*}

We report a tentative detection of faint \hi\ emission outside optical
body of NGC 7626.  Figure~\ref{fig:momnt0} presents the integrated
\hi\ emission contoured upon the DSS image, including both NGC 7626 
(east) and NGC 7619 (west). Figure~\ref{fig:n7626ch0} shows a
greyscale image of the DSS with the integrated \hi\ emission as thick
contours along with thinner contours for the radio continuum
emission. Figure~\ref{fig:n7626ch} shows the channel maps over a field
containing both NGC 7626 and NGC 7619. Since the single beam, single
channel flux is weak (with a peak line flux of 2 mJy beam$^{-1}$ or
4.3 sigma), the channel maps do not convincingly show the \hi\
emission. We therefore present two other figures:
Figure~\ref{fig:n7626pv} showing a position-velocity profile through
the peak emission and parallel to the right ascension axis, and
Figure~\ref{fig:n7626sp} showing a spectra extracted over the region
containing the two features in the moment map. The optical velocity of
NGC 7626 is indicated in both plots (by a dashed line in
Fig.~\ref{fig:n7626pv} and by an arrow in Fig.~\ref{fig:n7626sp}). 
While the emission is of low signal-to-noise, it appears in several 
adjacent channels. Still, the detection is considered tenative and 
should be confirmed by additional observations. 

The neutral hydrogen emission lies between NGC~7626 and NGC 7619 in
both space and velocity: the emission peak is at 23:20:32.6 +08:12:00
(J2000) and $V_\odot$=3518 km s$^{-1}$. This is 2.7$'$ (36 kpc) and
+85 km s$^{-1}$ from the optical center of NGC 7626
(Table~\ref{tab:results}), and 4.4$'$ (59 kpc) and -240 km s$^{-1}$
from NGC 7619. The flux integrated over the spectra of
Fig.~\ref{fig:n7626sp} is 0.17 Jy km s$^{-1}$, which is consistent
with the single dish upper limits mentioned above.  This flux
corresponds to an \hi\ mass of $8.3\times 10^7 M_\odot$ at the 
distance of NGC 7626. Curiously, the \hi\ cloud lies just to the side
of the radio jet (Fig.~\ref{fig:n7626ch0}). A very similar situation 
occurs in the shell Galaxy Cen A (NGC 5128; Schiminovich et al. 1994).

\begin{figure*}
\plotone{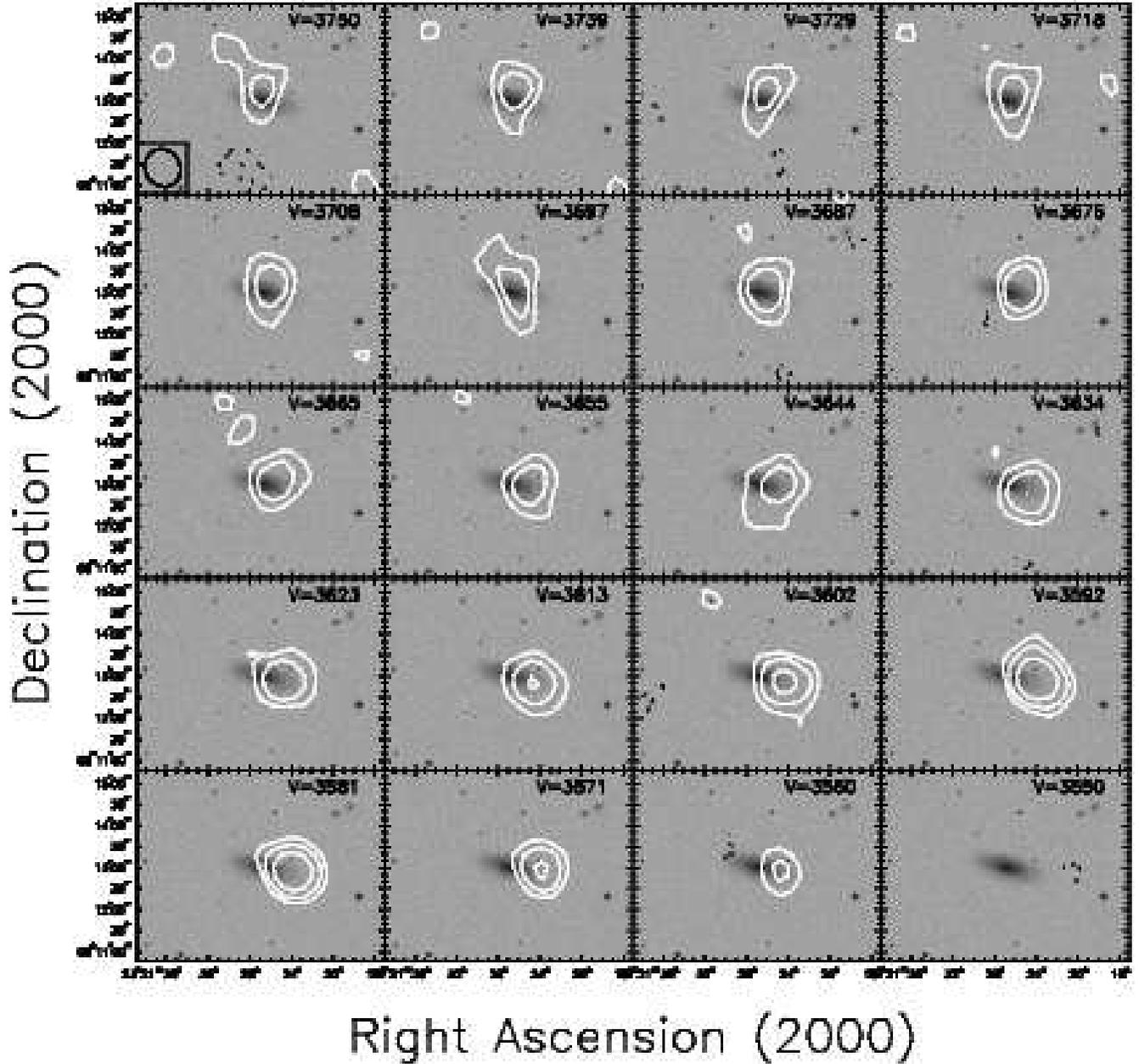}
\caption{
\hi\ contours upon optical image from the DSS for 
\hi\ channel maps contoured upon a greyscale representation of the 
DSS optical image for NGC 7631, the companion lying
11$'$ (145 kpc) to the east to NGC 7626.
The heliocentric velocity of each channel is given in the upper 
right of each panel, and the size of the synthesized beam
($55''\times 49''$ FWHM) is indicated by the inscribed ellipse
in the lower left corner of the first panel.
Contours are drawn at levels of [-3, 3, 6, 12] $\times$ 
0.64 mJy beam$^{-1}$, where the lowest contour corresponds to 
an \hi\ column density of $8.2\times 10^{18}$ cm$^{-2}$.
}
\label{fig:n7631ch}
\end{figure*}

We detect \hi\ in one other cluster member, the Sb galaxy NGC 7631
lying $11'$ (145 kpc) to east. This emission appears at the edge of
our bandpass, so we do not map the full extent of \hi\ line emission,
and as such the \hi\ flux and mass reported in Table~\ref{tab:results} 
are lower limits. Channel maps for this galaxy are presented in 
Figure~\ref{fig:n7631ch}. The \hi\ kinematics are characteristic of 
half of a rotating disk, with the higher velocity half falling outside
of our observing band. Three other cluster members fall within the
primary beam and velocity coverage of our \hi\ observations, but are 
not detected (see Table~\ref{tab:results}). 

\section{Discussion}
\label{sec:disc}

\subsection{\hi\ in the NGC 7626 System}

The detection of \hi\ outside the optical body of NGC 7626 was 
something of a surprise. However, our earlier study on the global
properties of the SS92 sample turned up a few shell systems with 
\hi\ located outside their optical bodies (Table 2 of Sansom et al. 
2000). Maps of these systems and other interesting examples appear in
the ``\hi\ Rogues Gallery'' (Hibbard et al. 2001a), in the category
``Peculiar Early Types with \hi\ Outside their Optical Body''
(Nos. 112--120 in the Rogues Gallery). This category is arranged in a
sequence ranging from systems with abundant quantities of \hi\ in the
outer regions to systems with little or no \hi\ in the outer regions,
with NGC 7626 appearing near the end of the sequence.

There are a number of systems in this sequence which are quite similar
to NGC 7626. Some specific examples are NGC 470/4, NGC 1316, NGC 4125,
and M$\,$86. NGC 474 is a spectacular shell system in the sample of
SS92, with \fsi\ = 5.26 and a heuristic merger age of 4--5 Gyr. There
are \hi\ clouds that appear to have been torn off the nearby Sb galaxy
NGC 470 and distributed at several locations around the galaxy, but
the \hi\ does not appear to make it into the optical body of the
galaxy (Schiminovich et al. 2001). Similarly, NGC 1316 (Fornax A) is a
well known peculiar system with many shells, ripples and plumes with
good evidence that it is a $\la$3 Gyr merger remnant (Schweizer 1980,
Goudfrooij et al. 2000a,b). There are \hi\ clumps around it that may
have been torn from the nearby S0/a galaxy NGC 1317, but again these
clouds avoid the main optical body of NGC 1316 (Horellou et
al. 2001). NGC 4125 is a shell galaxy in the sample of SS92, with
\fsi\ = 6.00 and a heuristic merger age of 6--8 Gyr. Unlike the other
systems, it is apparently isolated. It has a few small \hi\ clouds
outside its optical body (Rupen, Hibbard \& Bunker 2001). M$\,$86 (
a.k.a. NGC 4406) is a large early type galaxy in the Virgo cluster. It has an
X-ray ``tail'' to the northwest and an \hi\ cloud to the south,
outside the main body of M$\,$86 (Li \& van Gorkom 2001). NGC 7626 may
also bear some resemblance to the radio galaxy Coma A, where \hi\ has
been detected in absorption against the radio lobes, some $\sim$30 kpc
from the galaxy center (Morganti et al. 2002).

While there are systems similar to NGC 7626, the nature of the
intergalactic \hi\ clouds is not at all clear. They could be diffuse
remnant tidal debris which only remains neutral far outside of the body of
the elliptical, becoming photoionized by the stellar population at
smaller radii (Hibbard, Vacca \& Yun 2000); or they could be a result
of complex multiphase hydrodynamics, such as a jet-ISM, jet-ICM, or 
galaxy-ICM interaction. In this respect, X-ray imaging of the Pegasus 
group would be very instructive.

\subsection{\hi\ in fine-structure ellipticals}

We expected a fair detection rate in these high \fsi\
ellipticals. This presumption was based primarily on two
considerations: (1) peculiar E/SO galaxies have a 45\% \hi\ detection
rate (Bregman, Hogg, \& Roberts 1992; van Gorkom \& Schiminovich
1997), and (2) the expectation that the highest \fsi\ systems would be
the youngest remnants of disk mergers, and therefore most likely to
still have gas-rich tidal debris. Therefore, we were somewhat
surprised that only one of the five fine-structure galaxies that we
targeted was detected in \hi. Moreover, the detected system, NGC
7626, is on the low end of the \fsi\ scale, and we failed to detect
\hi\ in the three highest \fsi\ systems in the compilation of SS92 
(NGC 3610, NGC 3640 and NGC 4384).  

While the number of systems that we observed in the present program 
is much too small to draw any statistical conclusions, in our previous 
study (Sansom, Hibbard \& Schweizer 2000) we examined all of the available 
\hi\ mapping data for galaxies in the lists of SS92. We found that 17/37 
(46\%$\pm$16\%) were detected in \hi\ (i.e., the same fraction as
found by Bregman et al.\ for peculiar E/S0s). The \hi\ morphology
suggests a tidal origin in all but three of these detections.
However, the SS92 compilation includes early types with a wide range
of \fsi. When plotting cold and hot gas properties versus \fsi, we
find no correlation besides the previously known tendency for recent
($\la$3 Gyr) merger remnants to be X-ray faint (Fabbiano \& Schweizer
1995; Georgakakis, Forbes \& Norris 2000; O'Sullivan, Forbes \& Ponman
2001, see also http://www.star.uclan.ac.uk/$\sim$aes/GALS/ISM.html).

One possibility is that tidal \hi\ does not survive for as long in the
outer regions as inferred from studies of young, more isolated merger
remnants (Hibbard \& Mihos 1995). Another possibility is that 
a high fine structure index does not unambiguously identify recent 
disk-disk merger remnants. We look at each of these possibilities 
in turn.

Regarding the longevity of tidal material, in their dynamic study of
the merger remnant NGC 7252, Hibbard \& Mihos (1995) found that the
outermost \hi-rich material should remain at large radii from the
remnant, evolving on timescales of several Gyr. However, that study
only considered the effects of gravity, and did not address the
survivability of the gas in its neutral form. As the tidal debris
evolves, it becomes more diffuse. At low \hi\ column densities, this
gas may be susceptible to ionization by the intergalactic UV field
(Hill 1974, Bland-Hawthorn \& Maloney 1999), or even by the stellar
population of the remnant (Hibbard, Vacca \& Yun 2000).  Perhaps more
importantly, NGC 7252 is a relatively isolated system, while the five
systems studied here are in groups or clusters. In such environments,
the crossing times are of the same order as the timescale for tidal
evolution, and it may be difficult for tidal \hi\ to maintain a
coherent morphology for very long (see also Verdes-Montenegro et
al. 2001). We believe these effects contribute to the lack of
detectable tidal \hi\ in the targeted galaxies.  In this case, such
gas should give rise to extensive ionized haloes around galaxies
(Morris \& van den Bergh 1994, Tadhunter et al. 2000), which may be
detectable in absorption against background sources (e.g. Norman et
al. 1996, Carilli \& van Gorkom 1993). If we could accurately date the
time since the merger, it would help evaluate the likelihood of this
posibility.

Regarding the ability of a high fine-structure index to pick out the
youngest merger remnant, we re-evaluate the relationship between
fine-structure index and population age indicators\footnote{Here we 
make the usual assumption that disk-disk mergers must be
accompanied by a significant burst of star formation, lowering the
luminosity weighted mean age of the central stellar population. This is
a gross simplification, since merger-induced star formation histories 
are likely quite complex. The derived ``ages'' should considered as
age indicators, rather than actual ages.} derived by SS92.
The heuristic merger ages tabulated by SS92 were
based on broad band $UBV$ colors, which are not straight forward to
interpret in terms of ages, because of age-metallicity degeneracies,
non-solar abundances and ratios, line emission, dust, frosting of a
young population, etc. (see Worthey 1994 and Proctor \& Sansom 2002
for discussions of some of these effects). Spectroscopic
ages are now available using several line indicies sensitive to both
age and metallicity. We therefore re-evaluate the relationship between
fine-structure index and population age using these new age
indicators. 

\begin{figure*}
\plotone{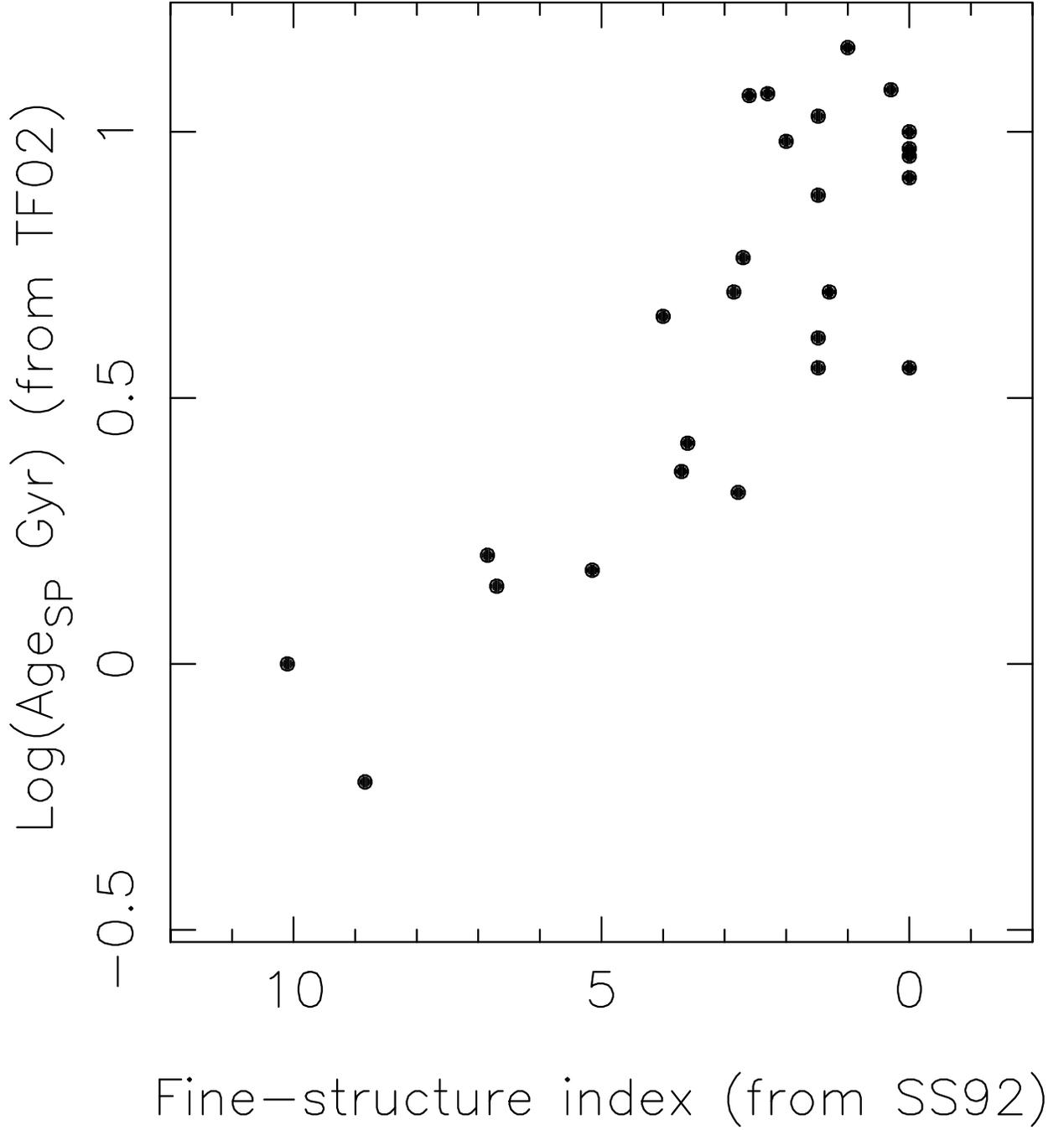}
\caption{The logarithm of the spectroscopically determined age 
from TF02 (Age$_{SP}$), derived using SSP fits to H$_{\beta}$ and
[MgFe] (combination of three) indices, versus the morphological
fine-structure index from SS92 ($\Sigma$), for the 26 galaxies in
common between the two samples.}
\label{fig:SigAge}
\end{figure*}

For our age indicators, we use the sample of early type galaxies with
uniform quality line-strength data assembled from the literature by
TF02, for which luminosity weighted ages have been determined using
the SSP models of Worthey (1994)\footnote{We note that the TF02 line
indicies have been converted to measurements within the inner $R_e$/8
of the systems, where $R_e$ is the effective radius of the galaxy, and
therefore are most sensitive to the ages of the central stellar
population.}. We emphasize that these are not absolute ages, but age
indicators: derived ages will differ for different models, but the
relative age ranking should be be maintained (see TF02 for a
discussion of these points). We therefore do not augment this sample
by including galaxies with merger ages estimated by a different
method. Figure \ref{fig:SigAge} shows how the fine-structure index
compares with the logarithm of the spectroscopically determined ages
(log [Age$_{SP}$]) for the 26 galaxies in the TF02 sample which also
have fine-structure quantified by SS92.

A statistical test on \fsi\ and Age$_{SP}$ for these 26 systems shows
that they are highly anti-correlated (Spearman rank-order correlation
coefficient $r_s$=-0.70). However, the correlation drops when the
young merger remnants NGC 3921 and NGC 7252 are dropped from the
sample ($r_s$=-0.62). Unlike the other systems in the SS92 sample,
these two systems were included based on their status as known merger
remnants. Further analysis indicates that most of the anti-correlation
between \fsi\ and Age$_{SP}$ comes from systems with high \fsi\
($\ga$4) and young ages ($\la$3 Gyr; $r_s$=-0.83); the rank order
correlation coefficient drops to $r_s$=-0.14 for systems with lower
\fsi\ and higher Age$_{SP}$.  Therefore, the apparent anti-correlation
may be due to the purposeful inclusion of the well-known merger
remnants NGC 7252 and NGC 3921 by both SS92 and TF02 at the high-\fsi,
low spectroscopic age end, and to the very few systems with a
fine-structure index in the range $6\la \Sigma \la 8$.  Indeed, three
of our systems fall within the range: NGC 3610, NGC 3640, NGC 4382
(Table~\ref{tab:VLAobs}). NGC 4382 is one of the points in the plot,
while NGC 3610 and NGC 3640 would lie well above the apparent
relationship if the age indicators given in \S\ref{sec:results} are 
taken (i.e., 4--7 Gyr for NGC 3610, and 6--8 Gyr for NGC 3640). So
while Fig.~\ref{fig:SigAge} reinforces the use of fine structure for
tracking dynamically young remnants, it would be more reassuring to
have more objects with a high fine-structure, and to have better age
indicators than luminosity weighted mean ages.

In conclusion, we cannot say unambiguously that the five
fine-structure ellipticals studied here are ``King Gap'' objects
(i.e., evolved remnants of major disk-disk mergers). They may indeed
be the result of gas-rich mergers, but evolved enough that any tidal
gas that was thrown out during the encounter has fallen back,
dispersed, or been converted to other forms. However, they may have
had quite different histories, including minor mergers or major
mergers of gas-poor progenitors.  Together, these results suggest that
progenitor type, encounter geometry, and local environment may all
play an important role in the expected post-merger evolution, and that
there is not one simple path from disk mergers to old ellipticals (see
also Hibbard et al. 2001a; Verdes-Montenegro et al. 2001). These
possibilities can only be constrained by investigating young objects
in the so-called ``King Gap''. To this end, Fig.~\ref{fig:SigAge}
offers some hope. Clearly, interacting systems may evolve in many ways
in this plane, depending on what merged and how many stars, if any,
were made during the interaction. Less drastic interacting histories
may leave remnants anywhere along the right hand side of this plot,
making it harder to determine the origin of such systems. However,
major mergers of gas-rich systems {\it must} evolve from the lower
left of the plot to the upper right.

It is not clear {\it how} remnants will evolve in this plane, or
whether \fsi\ and spectroscopically determined SSP ages are the best
parameters to evaluate: \fsi\ may not properly capture the dynamical
evolution of the optical peculiarities and, given the range of
possible progenitors and encounter geometries, perhaps no single
parameter can; additionally, mergers of gas rich systems may have
complicated star formation histories (e.g. Mihos \& Hernquist 1996)
that are poorly constrained by simple stellar population models (see
also Liu \& Green 1996), and luminosity weighted mean population ages
may only be roughly related to the 
time since the latest merger event (Hibbard et al. 2001b).  But the
above considerations reinforce the suggestion of SS92 that signs of
both dynamical youth and a youthful population should be used to
identify the youngest ``King Gap'' objects.

\section{Summary}
\label{sec:summary}

We detected neutral hydrogen in the vicinity of four of the five
fine-structure E/SO galaxies which we observed with the VLA
D-array. In only one case is the \hi\ directly associated with the
targeted elliptical. For this one case, NGC 7626, 
there is a tentative detection of \hi\ outside
the optical body of the galaxy. This system contains a kinematically
distinct core, but no obvious signs of being a recent merger, and its
stellar population is uniformly old (at least as indicated by SSP
determined ages). Therefore the origin of the \hi\ is unclear, but
similar systems are known from the literature.

Combining the present results with the analysis of Sansom, Hibbard \&
Schweizer 2000, we find that galaxies with a large value of the
fine-structure parameter, \fsi, are no more likely to be detected in
\hi\ than those with a low value of \fsi. This suggests that if the high
\fsi\ ellipticals are aged remnants of disk-disk mergers, their tidal
\hi\ did not survive longer than a few Gyrs. This may be due to the short
crossing time in the group environment, and/or due to ionization of
the \hi\ as it becomes more diffuse. This gas would contributed to a
very diffuse neutral or ionized halo around the remnants.

Given these results, we conclude that the most promising way to
constrain the observational characteristics of aged remnants of {\it
gas-rich} mergers would be to target younger ``King Gap'' objects
(ages $\la$2 Gyr) which contain signatures of both a dynamical youth
and a youthful central stellar population. A growing population of
objects is being identified, both by the appearance of optical and
gaseous tidal features (e.g. Schiminovich et al.\ 2001, Chang et
al. 2001, Hibbard et al. 2001a), and by spectroscopic studies of
peculiar or normal early type galaxies (e.g. Bergval, Ronnback \&
Johansson 1989; Oegerle, Hill \& Hoessel 1991; Zabludoff et al. 1996;
Longhetti et al. 1998, 2000; Hau et al. 1999; Trager et al. 2000; 
Georgakakis et al. 2001; TF02; Proctor \& Sansom 2002).

Once a population of recent merger remnants is identified, they would
enable us to judge the appearance of merger remnants at ages of $\ga$2
Gyr. By examining the cold gas content and distribution, global colors
and line indicies, inner/outer color differences, and light profiles
of such systems, we could evaluate whether they are likely to evolve
into normal ellipticals, or if instead they will leave long-lived
signatures of their merger origin (Mihos \& Hernquist 1994, Hibbard \&
Yun 1999, van den Marel \& Zurek 2000, Mihos 2001). By identifying the
expected observational signatures along the post-merger path we may
ultimately be able to determine the fraction of the elliptical galaxy
population which had a merger origin.

\acknowledgments

Our thanks to the staff at the VLA for their help in obtaining these
observations and for help during our preliminary analysis.  We thank
F. Schweizer, J. van Gorkom and D. Schiminovich for many fruitful
conversations and for comments on earlier drafts of this paper.

The VLA of the National Radio Astronomy Observatory is operated 
by Associated Universities, Inc., under
cooperative agreement with the National Science Foundation.
This research has made use of the NASA/IPAC Extragalactic Database
(NED), which is operated by the Jet Propulsion Laboratory, California
Institute of Technology, under contract with the National Aeronautics
and Space Administration. Optical images were taken from the Digitized
Sky Surveys, produced at the Space Telescope Science Institute under
U.S.  Government grant NAG W-2166. The images of these surveys are
based on photographic data obtained using the Oschin Schmidt Telescope
on Palomar Mountain.

\vfill\eject
\begin{deluxetable}{lccccc}
\tablecaption{Sample Properties}
\tablehead{
\colhead{Parameter} &
\colhead{NGC 3610} &
\colhead{NGC 3640} &
\colhead{NGC 4382} &
\colhead{NGC 5322} &
\colhead{NGC 7626} \\
}\startdata
Hubble Type (RC3) & E5: & E3 & SA(s)0+pec & E3--4 & E1 pec \\
Optical Velocity$^a$ (km s$^{-1}$)& 1696 & 1314 & 729 & 1781 & 3433 \\
Distance$^b$ (Mpc) & 29.2 & 24.2 & 16.8 & 31.6 & 45.6 \\
Fine-structure index$^c$ (\fsi) & 7.6 & 6.85 & 6.85 & 2.00 & 2.60 \\
Environment$^d$ & LGG 234 & LGG 233 & LGG 292 & LGG 360 & LGG 473 \\
& (N3642 group) & (N3640 group) & (Virgo I group) & 
(N5322 group) & (Pegasus group) \\
Known group members$^d$ & N=5 & N=7 & N=126 & N=10 & N=25 \\
\\
Observed Date & 1997 Dec 3 & 1997 Dec 3  & 1997 Dec 3  & 1997 Dec 3  & 1997 Dec 7  \\
Velocity Center$^e$ (km s$^{-1}$) & 1787 & 1314 & 760 & 1915 & 3423 \\
Velocity Coverage$^f$ (km s$^{-1}$) & 1516--2090 & 1044--1616 & 491--1061
& 1644--2218 & 3159--3729 \\
Phase Center:\\
~~~~~RA (J2000) & 11 18 25.9 & 11 21 06.8 & 12 25 24.7 & 13 49 15.6 & 23 20 42.4 \\
~~~~~DEC (J2000) & +58 47 14 & +03 14 08  & +18 11 27  & +60 11 29  & +08 13 02  \\
Time on Source & 2h 32m  & 2h 22m  & 1h 57m  & 2h 22m  & 2h 11m  \\
Flux Calibrator & 1331+305 & 1331+305 & 1331+305 & 1331+305 & 0137+331 \\
Phase Calibrator & 1206+642 & 1120+143 & 1254+116 & 1411+522 & 2253+161 \\
Bandpass Calibrator & 1206+642 & 1120+143 & 1331+305 & 1411+522 & 2253+161 \\
Bandwidth (MHz) & 3.25 & 3.25 & 3.25 & 3.25 & 3.25 \\
Number of Channels & 63 & 63 & 63 & 63 & 63 \\
Channel Spacing (km s$^{-1}$) & 10.4 & 10.4 & 10.4 & 10.4 & 10.5 \\
Synthesized Beam \\
~~~(major $\times$ minor axis) & 
$66'' \times 49''$ & 
$62'' \times 49''$ & 
$54'' \times 50''$ & 
$57'' \times 50''$ & 
$55'' \times 49''$ \\
rms noise (mJy beam$^{-1}$) & 0.60 & 0.63 & 0.65 & 0.58 & 0.46 \\
\enddata
\tablenotetext{a}{Optical velocities for NGC 3610 and NGC 5322 are 
from the Updated Zwicky Catalog (Falco et al.\ 1999), for NGC 4382 and
NGC 7626 are from Smith et al.\ (2000), and from the RC3 for NGC
3640.}
\tablenotetext{b}{Distances taken from the {\it Nearby Galaxy 
Catalog} (Tully 1988), except for NGC 7626, where the optical 
velocity from the RC3 is used, assuming a Hubble constant of 
$H_o=75$ km s$^{-1}$ Mpc$^{-1}$.}
\tablenotetext{c}{Fine-structure index from Table 1 of SS92.}
\tablenotetext{d}{Environment taken from Loose Galaxy Group catalog 
(Garcia, 1993), number of known group members taken from NED.}
\tablenotetext{e}{The central velocity for the \hi\ observations 
were based on the best available optical velocities at the time 
of the observations, which were from the RC3 for all galaxies except
NGC 4382, which was taken from Binggeli, Sandage and Tammann (1985).}
\tablenotetext{f}{Velocity coverage from channels 3--55 of \hi\ datacube.}
\label{tab:VLAobs}
\end{deluxetable}
\vfill\eject

{
\scriptsize
\begin{deluxetable}{llcrrrrrccc}
\tablecaption{VLA \hi\ Observations$^a$}
\tablehead{
\colhead{(1)} &
\colhead{(2)} &
\colhead{(3)} &
\colhead{(4)} &
\colhead{(5)} &
\colhead{(6)} &
\colhead{(7)} &
\colhead{(8)} &
\colhead{(9)} &
\colhead{(10)} \\
\colhead{Galaxy} &
\colhead{Hubble Type} &
\colhead{Field} &
\colhead{$V_{opt}$} &
\colhead{$V_{HI}$} &
\colhead{$\Delta W$} &
\colhead{$dr$} &
\colhead{$\rho$} &
\colhead{$(S_{HI} \Delta v)$} &
\colhead{$M_{HI}$} \\
 & & & (km s$^{-1}$) & (km s$^{-1}$) & (km s$^{-1}$) & 
(arcmin) & (kpc) & (Jy km s$^{-1}$) & ($M_\odot$) \\
}\startdata
NGC 3610     & E5: & NGC 3610 & 1696 &  --- & --- &    0 &   0 & $<$0.04 
& $< 2\times 10^7$ \\ 
\\
NGC 3640     & E3  & NGC 3640 & 1314 &  --- & --- &    0 &   0 & $<$0.05 
& $< 1\times 10^7$ \\ 
N3640 comp$^b$ & uncataloged
                   & NGC 3640 &  --- & 1180 &  62 & 15.1 & 105 & 2.4
&  $3.3\times 10^8$ \\
\\
NGC 4382  & S0+pec & NGC 4382 &  729 &  --- & --- &    0 &   0 & $<$0.06 
& $< 1\times 10^7$ \\ 
VCC 0797     & E?  & NGC 4382 &  773 &  --- & --- &  2.9 &  14 & $<$0.06 
& $< 1\times 10^7$ \\ 
NGC 4394     & SBb & NGC 4382 &  922 &  915 & 165 &  7.6 &  37 &   7.2
& $4.8\times 10^8$ \\
IC 3292     & dS0  & NGC 4382 &  710 &  --- & --- &  8.5 &  42 & $<$0.07
& $< 1\times 10^7$ \\ 
\\
NGC 5322   & E3--4 & NGC 5322 & 1781 &  --- & --- &    0 &   0 & $<$0.04 
& $< 2\times 10^7$ \\ 
MCG+10-20-039& unclassified 
                   & NGC 5322 &  --- & 1870 & 100 & 10.9 & 100 & 0.43
& $1.0\times 10^8$ \\
UGC 8714     & Im  & NGC 5322 & 2044 & 2030 & 210 & 23.2 & 215 & 4.7
& $1.1\times 10^9$ \\
\\
NGC 7626  & E1 pec & NGC 7626 & 3433 & 3518 & 105 &  2.7 &  36 & 0.17 
& $8.3\times 10^7$ \\
AGC 330257 & unclassified 
                   & NGC 7626 & 3470 &  --- & --- &  3.9 &  52 & $<$0.04
& $< 3\times 10^7$ \\  
NGC 7631$^c$ & Sb  & NGC 7626 & 3754 & $>$3700 & $>$200 & 11.0 & 145 
& $>$1.9 & $>9.3\times 10^8$ \\
NGC 7623 & S0$^0$: & NGC 7626 & 3739 &  --- & --- & 11.1 & 145 & $<$0.05
& $<5\times 10^8$ \\
UGC 12510 &   E    & NGC 7626 & 3542 &  --- & --- & 15.8 & 210 & $<$0.08
& $<8\times 10^8$ \\
\enddata
\label{tab:results}
\tablenotetext{(4)}{Optical radial velocity from NED}
\tablenotetext{(5)}{Velocity centroid of \hi\ line.}
\tablenotetext{(6)}{Separation from targeted elliptical in arcmin}
\tablenotetext{(7)}{Separation from targeted elliptical, assuming
both systems lie at the distance given in Table~\ref{tab:VLAobs}}
\tablenotetext{(8)}{Velocity width, calculated from intensity weighted
velocity map, and with no correction for inclination.}
\tablenotetext{(9) \& (10)}{6$\sigma$ limits, calculated from \hi\ 
datacube smoothed to a velocity resolution of 42 km s$^{-1}$ and
corrected for the primary beam attenuation.}
\tablenotetext{a}{Includes all \hi\ detections as well as
limits for all galaxies with known redshifts found 
in NED that fall within the primary beam ($15'$ radius) 
and velocity coverage of the VLA observations}
\tablenotetext{b}{Uncataloged highly-inclined LSB companion at
11:21:51.5 +03:24:17 (J2000).}
\tablenotetext{c}{\hi\ emission appears at the edge of the velocity
coverage of the observations, so the high velocity range of the \hi\
emission is not measured.}
\end{deluxetable}
}

\clearpage
\bigskip

\bigskip

\bigskip

\bigskip

\bigskip

\bigskip

\bigskip

\bigskip

\bigskip
\bigskip

\bigskip

\bigskip

\bigskip

\end{document}